\documentclass[aip,reprint]{revtex4-1}

\usepackage{graphicx}
\usepackage{hyperref}

\usepackage{xcolor} 
\usepackage{amsmath} 
\usepackage[percent]{overpic} 
\usepackage{subcaption}  
\usepackage[normalem]{ulem} 

\newcommand{\vect}[1]{\boldsymbol{#1}} 
\newcommand{\matr}[1]{\mathrm{\boldsymbol{#1}}} 
\newcommand{\ie}[1]{ #1^{(i)}} 
\newcommand{\ar}[1]{ #1^{(e)}} 
\newcommand{\dd}[2]{\frac{\partial #1}{\partial #2}}
\newcommand{\pvec}[1]{\vect{#1}'}

\draft 

\begin{document}


\title{Small deformation theory for a magnetic droplet in a rotating field} 



\author{A. P. Stikuts}
\email{andris\_pavils.stikuts@lu.lv}
\affiliation{MMML Lab, Faculty of Physics and Mathematics and Optometry, University of Latvia, Jelgavas iela 3 - 014, Riga, LV-1004, Latvia}
\affiliation{Sorbonne Université, CNRS, Laboratoire PHENIX, 4 place Jussieu, case 51, F-75005 Paris, France}

\author{R. Perzynski}
\affiliation{Sorbonne Université, CNRS, Laboratoire PHENIX, 4 place Jussieu, case 51, F-75005 Paris, France}

\author{A. Cēbers}
\affiliation{MMML Lab, Faculty of Physics and Mathematics and Optometry, University of Latvia, Jelgavas iela 3 - 014, Riga, LV-1004, Latvia}


\date{\today}

\begin{abstract}
A three dimensional small deformation theory is developed to examine the motion of a magnetic droplet in a uniform rotating magnetic field. 
The equations describing the droplet's shape evolution are derived using two different approaches - a phenomenological equation for the tensor describing the anisotropy of the droplet, and the hydrodynamic solution using perturbation theory. 
We get a system of ordinary differential equations for the parameters describing the droplet's shape, which we further analyze for the particular case when the droplet's elongation is in the plane of the rotating field.
The qualitative behavior of this system is governed by a single dimensionless quantity $\tau\omega$ - the product of the characteristic relaxation time of small perturbations and the angular frequency of the rotating magnetic field. 
Values of $\tau\omega$ determine whether the droplet's equilibrium will be closer to an oblate or a prolate shape, as well as whether it's shape will undergo oscillations as it settles to this equilibrium.
We show that for small deformations, the droplet pseudo-rotates in the rotating magnetic field - its long axis follows the field, which is reminiscent of a rotation, nevertheless the torque exerted on the surrounding fluid is zero.
We compare the analytic results with a boundary element simulation to determine their accuracy and the limits of the small deformation theory.
\end{abstract}


\maketitle 

\section{Introduction}

Due to a combination of responsiveness to external magnetic fields and their deformability, magnetic droplets make an interesting material that has found many applications.
They are widely used in microfluidics \cite{al-hetlani_continuous_2019}, where an external magnetic field can be used for their formation, transport and sorting. 
Magnetic droplets have been used as microrobots that scale obstacles and transport cargo \cite{fan_reconfigurable_2020, fan_ferrofluid_2020}.
Using time varying external fields, it is possible to induce dynamic self-assembled structures in magnetic droplet systems \cite{timonen_switchable_2013, wang_collective_2019, stikuts_spontaneous_2020}.
Biologically compatible magnetic droplets have found applications in biomedical context, such as a proposed method for treating retinal detachment \cite{mefford_field-induced_2007} and the measurement of mechanical properties of growing tissues \cite{serwane_vivo_2017}.

The behavior of magnetic fluid droplets in constant magnetic field has been widely researched. 
Such droplets elongate in the external field direction until the capillary forces balance out the magnetic forces. 
Assuming a spheroidal shape, theoretical equilibrium elongation depending on the magnetic field can be found \cite{bacri-instability-1982,tsebers-virial-1985,afkhami-deformation-2010}.
The development of the theoretical equilibrium curves allows for the experimental measurement of the droplet surface tension and magnetic permeability \cite{bacri-instability-1982,afkhami-deformation-2010,erdmanis-magnetic-2017}.
For large elongations, the droplets cease to be ellipsoidal and develop sharp tips \cite{stone-drops-1999} and numerical studies are required to describe their shape \cite{lavrova_numerical_2006, afkhami-deformation-2010, rowghanian-dynamics-2016}, which have shown that the ellipsoidal approximation is valid when the aspect ratio is $a/b\lesssim 4$\cite{rowghanian-dynamics-2016}.
Recently a semi-analytic relation has been proposed to describe the droplet elongation even in the large deformation regime
\cite{misra_magnetic_2020}.
The dynamics of a droplet in constant field can also be theoretically described with the ellipsoidal approximation \cite{tsebers-virial-1985,bacri_dynamics_1983}.

In rotating magnetic field, magnetic droplets show a large variety of complex behavior depending on the magnetic field parameters \cite{bacri-behavior-1994, janiaud_spinning_2000}.
For low frequencies, the droplets elongate and follow the field; viscous friction can cause the droplets to break \cite{sandre-assembly-1999}.
For high frequencies (much larger than the reciprocal of the characteristic time for surface deformations), the magnetic field forcing can be averaged over the rotation period. 
In this regime the droplet takes up interesting equilibrium shapes. 
At low magnetic field strengths the droplets are oblate spheroids flattened in the plane of rotating field; increasing the field strength, a spontaneous symmetry breaking occurs and they elongate tangential to the plane of rotation; increasing the field further, the droplets become flattened again in the rotation plane, and a crown of fingers can be seen on their perimeter \cite{bacri-behavior-1994}.
This oblate-prolate-oblate transition has been described theoretically assuming spheroidal shapes \cite{bacri-behavior-1994}, later it was extended to the case of an ellipsoid with 3 different axes \cite{morozov_bifurcations_2000}.
Recently an algorithm based on boundary element methods was used to calculate the equilibrium shapes in high-frequency fields without the constraint of ellipsoidal approximation \cite{erdmanis-magnetic-2017}. 
However, the time-dependent dynamics of magnetic droplets in rotating fields has not been yet thoroughly theoretically investigated. 

An approach often used to calculate droplet dynamics in external flows are phenomenological models for the anisotropy tensor, which describes the droplet's shape.
There are several definitions for the tensor quantity.
For example, models based on Doi-Ohta theory \cite{doi_dynamics_1991} up to a constant factor and summand use $\matr{q}=\int_S \vect{n}\vect{n} dS$, where $\vect n$ is the droplet's normal vector and the integral is over the droplet's surface \cite{almusallam_constitutive_2000}.
The Maffetone-Minale model \cite{maffettone_equation_1998} and others based on it describe the evolution of a tensor quantity $\matr{S}$, whose eigenvalues are the squared semiaxes of the ellipsoidal droplet.
Such phenomenological models have proven useful in predicting the experimentally observed droplet shapes in external flows \cite{boonen_droplet_2010}.
For a more thorough description of different models for an ellipsoidal droplet we refer to a review article \cite{minale_models_2010}.
The phenomenological nature of the models allows for their extension to more complicated situations, for example, non-newtonian fluids, taking into account wall effects and in our case - magnetic interaction.
It is important, however, to verify these models against experiments, microscopic theories and simulations.

In this paper, our goal is to derive analytic formulas for droplet shape dynamics in weak rotating magnetic fields, well below the measured threshold of shape instabilities.
We showcase a phenomenological model for droplet anisotropy tensor, and show how it can be used in case of a rotating (or precessing) field. 
By solving the full hydrodynamic problem, we show that its solution is exact for droplets with small deformations in weak fields. 
Furthermore, we explore the dynamics of magnetic droplets in rotating magnetic fields observing interesting effects, such as nonlinear shape oscillations and droplet pseudo-rotation without torque exertion.
We provide the velocity and pressure fields in and around the droplet valid for weak fields and small droplet deformations.
Last, we determine the applicability limits of the analytic expressions by comparison to numerical boundary element calculations.

\section{Problem formulation} \label{sec:problem_formulation}
We consider a superparamagnetic magnetic droplet with a magnetic permeability $\mu$, surface tension coefficient $\gamma$, viscosity $\ie\eta$ and volume $4\pi R_0^3/3  $ suspended in an infinite carrier fluid whose viscosity is $\ar\eta$ and magnetic permeability equal to the vacuum magnetic permeability $\mu_0$. A rotating external field is applied
$\vect{H}_\infty = H_\infty \left( \vect e_y \cos(\omega t) + \vect e_z \sin(\omega t) \right)$
with an angular velocity $\vect\omega = \omega \vect e_x$. $\vect H$ is the magnetic field intensity, $\vect B=\mu_0(\vect{H} + \vect M)$ is the magnetic field induction and $\vect M$ is the magnetization. Magnetization is assumed to be linearly dependent on the magnetic field intensity $\vect M = \chi \vect H$, and therefore so is the magnetic field induction $\vect B = \mu \vect H$, where the susceptibility is $\chi=\mu/\mu_0-1$.

The fluid both inside and outside the droplet obeys the Stokes equations
\begin{equation}
    -\dd{\tilde{p}}{x_i}+\eta \frac{\partial^2 v_i}{\partial x_j \partial x_j} + \dd{T_{ij}}{x_j} = 0,\ \dd{v_j}{x_j}=0,
\label{eq:stokes_eq}
\end{equation}
where $v_i$ is the fluid velocity and $\tilde{p}$ is its pressure. $T_{ij} = -\mu_0 H^2 \delta_{ij}/2+H_iB_j$ is the Maxwell stress tensor such that $\partial_j T_{ij}=\mu_0 M_j \partial_j H_i$ \cite{rosensweig-ferrohydrodynamics-2014,blums-magnetic-1997}. 

For linearly magnetizable fluids, its action in the bulk fluid can be represented by a magnetic pressure term
\begin{equation}
    p_M = -\frac{1}{2} \mu_0 \left( \frac{\mu}{\mu_0} - 1 \right)H^2,\  \dd{T_{ij}}{x_j} = -\dd{p_M}{x_i}.
\end{equation}
This allows us to formulate the Stokes equations of motion with an effective pressure $p=\tilde{p}+p_M$ taking into account $p_M$ in the boundary conditions.

They are as follows. On the droplet surface the velocity is continuous 
\begin{equation}
    \ar{\vect{v}}-\ie{\vect{v}} = 0.
\label{eq:v_BE}
\end{equation}
The total force on the surface element in the normal direction is zero
\begin{equation}
    n_i(\ar{\sigma}_{ij}-\ie{\sigma}_{ij})n_j + f_M - \gamma (k_1+k_2) = 0.
\label{eq:normal_force_BE}
\end{equation}
$\sigma_{ij}=-p\delta_{ij}+\eta\left( \partial v_i/ \partial x_j + \partial v_j / \partial x_i \right)$ is the effective hydrodynamic stress tensor (including the magnetic pressure), $n_i$ is the droplet's outward unit normal, $k_1$ and $k_2$ are the principal curvatures and $f_M$ is the effective magnetic surface force
\begin{equation}
\label{eq:mag_surf_force}
    f_M = \frac{1}{2} \mu_0 \left( \frac{\mu}{\mu_0} - 1 \right)\left( \frac{\mu}{\mu_0} {\ie{H}_n}^2 + {\ie{H}_t}^2  \right),
\end{equation}
where $H_n$ and $H_t$ are the normal and tangential field components, respectively, on the surface of the droplet. 
The total force in the tangential direction on the surface element is zero
\begin{equation}
    P_{ki}(\ar{\sigma}_{ij}-\ie{\sigma}_{ij})n_j=0,
\label{eq:tang_force_BE}
\end{equation}
where $P_{ki}=\delta_{ki}-n_k n_i$ is the projection operator on the tangent plane of the surface. 

To find the magnetic field, we solve the equations of magnetostatics
\begin{equation}
\label{eq:magnetostatics}
    \vect{\nabla}\times\vect{H} = 0, \ \vect{\nabla}\boldsymbol{\cdot}\vect{B} = 0.
\end{equation}
Introducing the magnetic potential $\psi$ such that $\vect{H}=\vect{\nabla}\psi$, the equations \eqref{eq:magnetostatics} are satisfied if it satisfies the Laplace equation
\begin{equation}
\label{eq:laplacePsi}
    \Delta \psi = 0.
\end{equation}

The boundary conditions for the magnetostatic problem are as follows. On the surface of the droplet the normal component of $\vect B$ is continuous
\begin{equation}
    \mu_0 n_i \dd{\ar{\psi}}{x_i} - \mu n_i \dd{\ie{\psi}}{x_i} = 0 ,
\end{equation}
and, if there is no distribution of magnetic dipoles on the surface, the potential itself is continuous
\begin{equation}
    \ar{\psi} - \ie{\psi} = 0,
\end{equation}
which also automatically satisfies the requirement that the tangential component of $\vect H$ is continuous.
Lastly, far away from the droplet
\begin{equation}
    \dd{\ar{\psi}}{x_i} = H_{\infty i}.
\end{equation}

\subsection{Dimensionless parameters}
Three dimensionless parameters naturally arise in the solution of the problem: 
\begin{itemize}
    \item the viscosity ratio $\lambda = \ie{\eta} / \ar{\eta}$,
    \item the relative magnetic permeability $\mu_r = \mu / \mu_0$,
    \item the magnetic Bond number $Bm=4\pi\mu_0 R_0 H_\infty^2 / \gamma$, which is the ratio of a characteristic magnetic force and a characteristic surface tension force.
\end{itemize}
It is worth mentioning that different authors define different dimensionless groups as the magnetic Bond number. 
For example, some definitions are without the $4
\pi$ factor \cite{ sandre-assembly-1999} $Bm=\mu_0 R_0 H_\infty^2 / \gamma$, others have a $2$ in the denominator \cite{afkhami-deformation-2010} $Bm=\mu_0 R_0 H_\infty^2 / (2\gamma)$. 
Whereas some call this ratio the magnetic capillary number $Ca_m$ \cite{cunha-effects-2020}, analogous to a similar quantity for droplets in electric fields - the electric capillary number $Ca_{el}$ \cite{vlahovska-rheology-2011}.
Our choice corresponds to the one used in the work of Erdmanis et al.\cite{erdmanis-magnetic-2017}
Care should be exercised when comparing results between them.

\subsection{Description of droplet's shape}

\begin{figure}[h]
    \centering
    \begin{overpic}[width=0.3\textwidth]{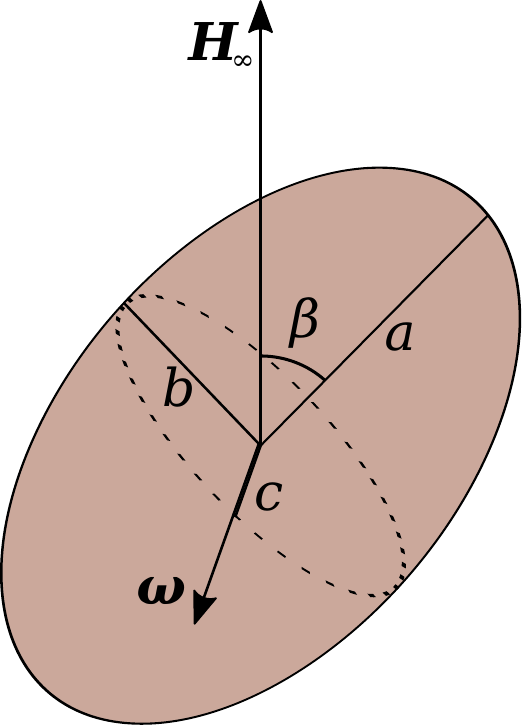}
        \put (-20,90) {\textcolor{black}{elongation: $\epsilon_1=\frac{a-b}{b}$}}
        \put (-20,70) {\textcolor{black}{flatness: $\epsilon_2=\frac{b-c}{b}$}}
    \end{overpic}
    \caption{In the limit of small deformations, the droplet can be described as an ellipsoid with semiaxes $a\geq b\geq c$ and with an angle $\beta$ between the longest axis and the field in the case when the motion is in the rotation plane of the magnetic field. $\beta<0$ if the droplet lags behind the field. The magnetic field $\vect{H}_\infty$ rotates with the angular velocity $\vect\omega$.
    }
    \label{fig:abc_beta_definitions}
\end{figure}

In the limit of weak field ($Bm \ll 1$) when the deformation of the magnetic droplet is small, we may represent it as a triaxial ellipsoid with semiaxes $a\geq b \geq c$ (figure \ref{fig:abc_beta_definitions}). 
Equivalently we can use two deformation parameters, the elongation
\begin{equation}
\label{eq:eps1_definition}
    \epsilon_1=\frac{a-b}{b},
\end{equation}
and the flatness
\begin{equation}
\label{eq:eps2_definition}
    \epsilon_2=\frac{b-c}{b},
\end{equation}
of the droplet, which together with the incompressibility condition $abc=R_0^3$ fully determine all three semiaxes. 
These deformation parameters are related to the commonly used \cite{vlahovska-rheology-2011, rallison-deformation-1984,das-nonlinear-2017, rowghanian-dynamics-2016} Taylor deformation parameter $D$ in the limit where the droplet is a prolate spheroid as
\begin{equation}
    D=\frac{\epsilon_1}{\epsilon_1+2} = \frac{\epsilon_1}{2} + O(\epsilon_1^2),
\end{equation}
and in the limit where the droplet is an oblate spheroid as
  \begin{equation}
    D=\frac{\epsilon_2}{\epsilon_2-2} = -\frac{\epsilon_2}{2} + O(\epsilon_2^2).
\end{equation}

When the droplet deforms in the plane of the rotating magnetic field, the smallest axis $c$ is in the direction of $\vect{\omega}$, and another parameter arises - the angle $\beta$ between the droplet's longest axis and the external magnetic field direction. 
It is chosen such that $\beta<0$ if the droplet's axis is lagging behind the field.
In general, two more angles would be necessary to describe the out-of-plane motion of the droplet, however, we do not examine it in this work.

\section{Anisotropy tensor approach}
\label{sec:anisotropy_tensor}
Ellipsoidal droplet's shape is described by a quadratic form
\begin{equation}
    x_i A_{ij} x_j = 1.
\end{equation}
Unperturbed spherical shape corresponds to $A_{ij}=\delta_{ij}/R_0^2$, where $\delta_{ij}$ is the Kronecker delta. For deformed droplets we may write $A_{ij}=(\delta_{ij}+\zeta_{ij})/R_0^2$, where $\zeta_{ij}$ is the symmetric anisotropy tensor. Eigenvalues of $\zeta_{ij}$ are $\lambda_i=(R_0^2/a_i^2 - 1)$, where $a_i$ are the semiaxes of the droplet.
For small deformations the eigenvalues can be written as
\begin{equation}
    \left\{ \begin{aligned} 
    \lambda_1 &= -2\left(\frac{c}{R_0}-1\right) \\
    \lambda_2 &= -2\left(\frac{b}{R_0}-1\right) \\
    \lambda_3 &= -2\left(\frac{a}{R_0}-1\right) 
    \end{aligned} \right. .
\label{eq:lambda_to_a}
\end{equation}
Conservation of volume requires $ \prod_{i=1}^3 (1+\lambda_i) = 1$. For small deformations, it means that 
\begin{equation}
    \lambda_1+\lambda_2+\lambda_3=0.
\label{eq:const_volume}
\end{equation}

Let us consider a rotating coordinate frame $(\pvec{e}_x,\pvec{e}_y,\pvec{e}_z)$ such that
\begin{equation}
\left\{ \begin{aligned} 
  \pvec{e}_x &= \vect{e}_x \\
  \pvec{e}_y &= \vect e_y \sin(\omega t) - \vect e_z \cos(\omega t) \\
  \pvec{e}_z &= \vect e_y \cos(\omega t) + \vect e_z \sin(\omega t)
\end{aligned} \right.
\end{equation}
In this rotating frame, the magnetic field is stationary and in the $\pvec{e}_z$ direction, and there is an additional constant fluid flow rotating with an angular velocity $\vect\Omega=-\omega\pvec{e}_{x}$. 

In static case when $\omega=0$, the droplet elongates in the $\pvec{e}_z$ direction to an axisymmetric equilibrium shape characterized by its elongation
\begin{equation}
    \delta=\left[\epsilon_1\right]_{\omega=0}=\left[\frac{a-b}{b}\right]_{\omega=0}
\end{equation}
In that case we can write the equilibrium anisotropy tensor
\begin{equation}
\matr{\zeta}^{0}=\begin{pmatrix}
\frac{2\delta}{3}&0&0\\
0&\frac{2\delta}{3}&0\\
0&0&-\frac{4\delta}{3}
\end{pmatrix}.
\end{equation}

A phenomenological equation for the tensor $\zeta_{ij}$ in the rotating frame reads \cite{dikanskii-magnetic-1990}
\begin{equation}
\frac{d\zeta_{ij}}{dt}-e_{ipr}\Omega_{p}\zeta_{rj}-e_{jpr}\Omega_{p}\zeta_{ir}=-\frac{1}{\tau}(\zeta_{ij}-\zeta^{0}_{ij})
\label{eq:zeta_eq}
\end{equation}
where $\tau
$ is the relaxation time of droplet's shape perturbations and $e_{ipr}$ is the Levi-Civita symbol. 
A brief justification of eq. \eqref{eq:zeta_eq} might be appropriate here.
The left-hand side of the equation is the Jaumann derivative, which takes into account that the components of the tensor can change due to rotation.
Whereas the right-hand side states that the droplet's shape exponentially approaches an equilibrium described by $\matr{\zeta^0}$ with a characteristic time $\tau$.
It might be noted that an additional term can be added to describe the effect of a shear flow on the droplet's shape and thus describe their rheology, but it is outside of the scope of this work. 
See, for example, the Maffettone-Minale model \cite{maffettone_equation_1998} for ellipsoidal droplets in a viscous flow, which uses a tensor $\matr S$ close to the inverse of the one used here $\matr{S}=R_0^2(\matr{\zeta}+\matr{I})^{-1}$, where $\matr{I}$ is the identity tensor.

For the components of the anisotropy tensor, the equation \eqref{eq:zeta_eq} gives three independent sets of equations
\begin{equation}
    \frac{d \zeta_{11}}{dt} = -\frac{1}{\tau}\left( \zeta_{11} -\frac{2\delta}{3} \right),
    \label{eq:1system}
\end{equation}
\begin{equation}
    \left\{ \begin{aligned} 
    \frac{d \zeta_{12}}{dt} &= -\frac{1}{\tau} \zeta_{12} + \omega \zeta_{13} \\
    \frac{d \zeta_{13}}{dt} &= -\frac{1}{\tau} \zeta_{13} - \omega \zeta_{12} 
    \end{aligned} \right. ,
    \label{eq:2system}
\end{equation}
\begin{equation}
    \left\{ \begin{aligned} 
    \frac{d \zeta_{22}}{dt} &= -\frac{1}{\tau}\left( \zeta_{22} -\frac{2\delta}{3} \right) + 2\omega \zeta_{23} \\
    \frac{d \zeta_{23}}{dt} &= -\frac{1}{\tau} \zeta_{23} - \omega \zeta_{22} + \omega \zeta_{33} \\
    \frac{d \zeta_{33}}{dt} &= -\frac{1}{\tau}\left( \zeta_{33} +\frac{4\delta}{3} \right) - 2\omega \zeta_{23}
    \end{aligned} \right. .
    \label{eq:3system}
\end{equation}

When the droplet moves in the plane of the magnetic field, only the system of equations \eqref{eq:3system} is relevant. 
The first equation \eqref{eq:1system} is just needed to satisfy the incompressibility condition for the droplet, it gives no new information since the trace of $\zeta_{ij}$ is zero as follows from eq. \eqref{eq:const_volume}. 
The equations \eqref{eq:2system} describe the out-of-plane motion of the droplet, their solutions decay exponentially to zero and therefore the in-plane motion is stable.

Since $\zeta_{ij}$ is symmetric, it has three real eigenvalues and mutually orthogonal eigenvectors $(\vect{n}^{1},\vect{n}^{2},\vect{n}^{3})$. We can write
\begin{equation}
\zeta_{ij}=\lambda_{1}n^{1}_{i}n^{1}_{j}+\lambda_{2}n^{2}_{i}n^{2}_{j}+\lambda_{3}n^{3}_{i}n^{3}_{j}.
\end{equation}
The eigenvectors are aligned with the axes of the ellipsoid, hence they are 
\begin{equation}
\left\{ \begin{aligned} 
  \vect{n}^1 &= \pvec{e}_x \\
  \vect{n}^2 &= \pvec{e}_y \cos(\beta) + \pvec{e}_z \sin(\beta) \\
  \vect{n}^3 &= -\pvec{e}_y \sin(\beta) + \pvec{e}_z \cos(\beta) 
\end{aligned} \right. .
\label{eq:eigenvectors}
\end{equation}
We then see that
\begin{equation}
\left\{ \begin{aligned} 
  \zeta_{22} &= \lambda_2\cos^2(\beta) + \lambda_3\sin^2(\beta)  \\
  \zeta_{23} &= (\lambda_2-\lambda_3)\cos(\beta)\sin(\beta)   \\
  \zeta_{33} &= \lambda_2\sin^2(\beta) + \lambda_3\cos^2(\beta) 
\end{aligned} \right. .
\label{eq:zetas}
\end{equation}
Inserting eq. \eqref{eq:zetas} in eq. \eqref{eq:3system}, we find the equations for the time evolution of the eigenvalues of the anisotropy tensor and the angle $\beta$
\begin{equation}
\left\{ \begin{aligned} 
  \frac{d \lambda_2}{d t} &= -\frac{1}{\tau} \left( \lambda_2 + \frac{\delta}{3} -\delta\cos(2\beta) \right) \\
  \frac{d \lambda_3}{d t} &= -\frac{1}{\tau} \left( \lambda_3 + \frac{\delta}{3} +\delta\cos(2\beta) \right) \\
  \frac{d \beta}{d t} &= -\omega + \frac{\delta \sin(2\beta)}{\tau(\lambda_3-\lambda_2)}
\end{aligned} \right. .
\label{eq:lambda_system}
\end{equation}

Finally, from eq. \eqref{eq:lambda_to_a} it follows that for small deformations, the droplet deformation parameters (defined in eq. \eqref{eq:eps1_definition} and eq. \eqref{eq:eps2_definition}) are connected with $\lambda_i$ by
\begin{equation}
\epsilon_1 = \frac{1}{2}(\lambda_2-\lambda_3), \quad \epsilon_2 = \frac{1}{2}(\lambda_1-\lambda_2).
\label{eq:lambda_to_eps}
\end{equation}
Combining eq. \eqref{eq:lambda_system}, eq. \eqref{eq:lambda_to_eps} and eq. \eqref{eq:const_volume}, we get the time evolution of the deformation parameters
\begin{equation}
\left\{ \begin{aligned} 
  \frac{d \epsilon_1}{d t} &= -\frac{1}{\tau} \left( \epsilon_1 - \delta \cos(2\beta) \right) \\
  \frac{d \epsilon_2}{d t} &= -\frac{1}{\tau} \left( \epsilon_2 - \delta \sin^2(\beta) \right) \\
  \frac{d \beta}{d t} &= -\omega - \frac{\delta \cos(\beta)\sin(\beta)}{\tau \epsilon_1}
\end{aligned} \right. .
\label{eq:eps_system}
\end{equation}
The equations \eqref{eq:zeta_eq} used to derive the system \eqref{eq:eps_system} can be considered phenomenological, where the constants $\delta$ and $\tau$ may be determined experimentally or on the basis of some microscopic model considered in the next section. 

Note that the same anisotropy tensor equation \eqref{eq:zeta_eq} can be used to calculate the droplet dynamics in a precessing field. One just has to change reference frame such that the field is stationary.

\section{Hydrodynamic approach}
\label{sec:hydrodanamic}
We solve the problem formulated in section \ref{sec:problem_formulation} asymptotically for small deformations using perturbation theory. The droplet shape is slightly deviated from a sphere, which is described in spherical coordinates
\begin{equation}
\left\{ \begin{aligned} 
    x &= r \cos(\phi)\sin(\theta) \\
    y &= r \sin(\phi)\sin(\theta) \\
    z &= r\cos(\phi)
\end{aligned} \right. ,
\end{equation}
by
\begin{equation}
    r=R_0\left(1+\varepsilon f(\theta, \phi)\right),
\label{eq:deformed_surface}
\end{equation}
where $\varepsilon$ is a small parameter, and we use $x,y,z$ to denote the coordinates in the basis of $\pvec{e}_x,\pvec{e}_y,\pvec{e}_z$.
We assume that all the physical parameters can be expressed as a power series of  $\varepsilon$ 
\begin{equation}
\begin{aligned}
    \vect{v} &= \vect{v}_0 + \varepsilon \vect{v}_1 + O(\varepsilon^2), \\
    p &= p_0 + \varepsilon p_1 + O(\varepsilon^2), \\ 
    f_M &= {f_M}_0 + \varepsilon {f_M}_1 + O(\varepsilon^2).
\end{aligned}
\end{equation} 
In the solution, we identify $\varepsilon$ with the elongation and flatness parameters $\epsilon_1=O(\varepsilon)$, $\epsilon_2=O(\varepsilon)$. 
Furthermore, similarly to how it was done in paper by Vlahovska \cite{vlahovska-rheology-2011}, we assume that the magnetic Bond number is small. 
Therefore we omit from the solution the terms which are $O(\varepsilon^2)$ and $O(\varepsilon Bm)$.

Variables with the index 0 are the solutions to the spherical droplet, they correspond to the flow arising from the magnetic forcing.
Next order corrections $\varepsilon \vect{v}_1, \varepsilon p_1$ arise from the effects of surface deformation, such as the flow due to surface tension.
$\vect{v}, p$ and $\vect{v}_0, p_0$ satisfy the Stokes equations (eq. \eqref{eq:stokes_eq}), therefore so do $\varepsilon \vect{v}_1, \varepsilon p_1$. 

First, we find the solution for a non-rotating magnetic field. Then to get the solution for a rotating field, instead of having the magnetic field rotate, we add a flow field rotating in the opposite direction.

\subsection{Solution for a spherical droplet}
The expression for the magnetic field inside a spherical magnetic (or equivalently dielectric) droplet placed in a homogeneous external field $\vect{H}_\infty=H_\infty\pvec{e}_z$ is well known \cite{stratton-electromagnetic-1941}
\begin{equation}
    \ie{\vect{H}}_0=\frac{3H_\infty}{\mu+2}\pvec{e}_z ,
\end{equation}
which leads to the effective magnetic surface force (eq. \eqref{eq:mag_surf_force}) of
\begin{equation}
\begin{split}
    {f_M}_0 
    =& \frac{\gamma}{R_0} \frac{9Bm}{8\pi}\frac{\mu_r-1}{(\mu_r+2)^2}\left( \mu_r \cos^2(\theta)+\sin^2(\theta) \right).
\end{split}
\end{equation}

To find the velocity, we use the Lamb's solution for the Stokes equations in spherical harmonics \cite{lamb_hydrodynamics_1975, kim-microhydrodynamics-1991}. For the flow outside and inside the droplet, it reads
\begin{equation}
\label{eq:Lambs_sol_ar}
\begin{split}
    \ar{\vect{v}} =& \sum_{l=1}^\infty \frac{1}{\ar{\eta}}\left[ -\frac{(l-2)r^2\vect{\nabla} \ar{p}_{-l-1}}{2l(2l-1)} + \frac{(l+1)\vect r \ar{p}_{-l-1}}{l(2l-1)}  \right] \\
    &+ \sum_{l=0}^\infty \left[ \vect{\nabla} \ar{\phi}_{-l-1} + \vect{\nabla}\times\left(\vect r \ar{\chi}_{-l-1}\right) \right],
\end{split}
\end{equation}
\begin{equation}
\label{eq:Lambs_sol_ie}
\begin{split}
    \ie{\vect{v}} =& \sum_{l=0}^\infty \frac{1}{\ie{\eta}}\left[ \frac{(l+3)r^2\vect{\nabla} \ie p_l}{2(l+1)(2l+3)} - \frac{l\vect r \ie p_l}{ (l+1)(2l+3)}  \right] \\
    &+ \sum_{l=0}^\infty \left[ \vect{\nabla} \ie{\phi}_l + \vect{\nabla}\times\left(\vect r \ie{\chi}_l\right) \right],
\end{split}
\end{equation}
and the corresponding effective pressure reads
\begin{equation}
    \ar{p}=\sum_{l=1}^\infty \ar{p}_{-l-1},\quad \ie{p}=\sum_{l=0}^\infty \ie{p}_{l}, 
\end{equation}

where $\ie p_l$, $\ar p_{-l-1}$, $\ie \phi_l$, $\ar \phi_{-l-1}$, $\ie \chi_l$ and $\ar \chi_{-l-1}$ are a sum of spherical harmonics of degree $l$ multiplied by an appropriate power of $r$: 
\begin{equation*}
    \ie p_l =  r^l\sum_{m=-l}^l a_l^m Y_l^m, \quad \ar p_{-l-1} =  r^{-l-1}\sum_{m=-l}^l A_l^m Y_l^m,
\end{equation*}
\begin{equation*}
    \ie \phi_l =  r^l\sum_{m=-l}^l b_l^m Y_l^m, \quad \ar \phi_{-l-1} =  r^{-l-1}\sum_{m=-l}^l B_l^m Y_l^m,
\end{equation*}
\begin{equation*}
    \ie \chi_l =  r^l\sum_{m=-l}^l c_l^m Y_l^m, \quad \ar \chi_{-l-1} =  r^{-l-1}\sum_{m=-l}^l C_l^m Y_l^m.
\end{equation*}
The coefficients $a,A,b,B,c,C$ are to be found from the boundary conditions. The spherical harmonics are 
\begin{equation}
    Y_l^m(\theta,\phi) = \sqrt{\frac{2l+1}{4\pi}\frac{(l-m)!}{(l+m)!}}P_l^m\left(\cos(\theta)\right)e^{i m \phi},
\end{equation}
where $P_l^m$ is the associated Legendre polynomial. It is worth noting that the spherical harmonics are orthonormal with respect to integration over the whole solid angle \cite{jackson-classical-1998}
\begin{equation}
\label{eq:spher_harm_orthon}
    \int_0^{2\pi} d\phi \int_0^{\pi} Y_l^m(\theta,\phi) \bar{Y}_{l'}^{m'}(\theta,\phi) \sin \theta d \theta=\delta_{ll'}\delta_{mm'},
\end{equation}
where the bar denotes the complex conjugate. Therefore, the coefficients in the spherical harmonic series of function $f(\theta,\phi)=\sum_{l=0}^{\infty}\sum_{m=-l}^{l}A_l^m Y_l^m(\theta,\phi)$ can be found by $A_l^m = \int_0^{2\pi} d\phi \int_0^{\pi} f(\theta,\phi) \bar{Y}_{l}^{m}(\theta,\phi) \sin \theta d \theta$.

The normal force on the spherical droplet boundary ($r=R_0$) that is due to magnetic field and surface tension is
\begin{equation}
\begin{aligned}
   f_M &- \gamma(k_1+k_2)  \\
   =&\frac{\gamma}{R_0}\left( \frac{9Bm}{8\pi}\frac{\mu_r-1}{(\mu_r+2)^2}\left( \mu_r \cos^2(\theta)+\sin^2(\theta) \right) -2 \right).
\end{aligned}
\end{equation}
It must be counteracted by an effective hydrodynamic stress such that eq. \eqref{eq:normal_force_BE} is satisfied.
Similarly to the work by Das and Saintillan\cite{das-three-dimensional-2021}, we can see which summands in the Lamb's solution (eq. \eqref{eq:Lambs_sol_ar}, eq. \eqref{eq:Lambs_sol_ie}) need to be retained to produce the effective hydrodynamic stress of the right form. 
The normal force contains only harmonics of degree $l=0$ and $l=2$, and there is a symmetry around $\phi$.
Therefore, the only non-zero coefficients in the Lamb's solution that are needed are $a_0^0$, $B_0^0$, $a_2^0$, $b_2^0$, $A_2^0$, $B_2^0$. 

Now that there is a finite number of unknown variables, the further calculations are straight forward, albeit cumbersome, therefore were done in Wolfram Mathematica. The Lamb's solution with the non-zero coefficients is substituted in the boundary equations (eq. \eqref{eq:v_BE},  eq. \eqref{eq:normal_force_BE}, eq. \eqref{eq:tang_force_BE}). We write the boundary conditions on the spherical drop ($r=R_0$), and expand them in spherical harmonics. For the boundary conditions to be satisfied for arbitrary angles $\theta$ and $\phi$, we require that each coefficient of expansion in spherical harmonics is zero, which leads to six equations for the six unknown coefficients in the Lamb's solution.

The resulting velocities and effective pressures for the case of a spherical droplet can be found in appendix \ref{sec:appendix_spherical}.

\subsection{Solution for a deformed droplet}
We assume that the droplet is ellipsoidal.
Such an assumption is justified since the magnetic force induces changes in shape of the droplet corresponding to spherical harmonic $Y_2^0$, which for small deformations means an ellipsoidal shape. 
And as can be seen further, in a rotational flow, initially ellipsoidal droplets remain ellipsoidal.
For small deformation parameters $\epsilon_1$ and $\epsilon_2$ in Cartesian coordinates, the equation for an ellipsoid, whose largest axis is rotated by an angle $\beta$ from $\pvec{e}_z$ (magnetic field direction) is
\begin{equation}
\begin{aligned}
   \frac{\epsilon_1-\epsilon_2}{3}+(1+2\epsilon_2)\frac{x^2}{R_0^2} +\frac{ \left(y\cos(\beta) + z\sin(\beta)\right)^2}{R_0^2}\\ +\left(1-2\epsilon_1\right)\frac{\left(z\cos(\beta) - y\sin(\beta)\right)^2}{R_0^2}=1.
\end{aligned}
\end{equation}
In spherical coordinates the equation becomes
\begin{equation}
 \begin{aligned} 
    r &= R_0\left(1 + \varepsilon f(\theta,\phi)\right) \\
    &= R_0\left(1 + \alpha_2^2 Y_2^{-2} + \alpha_2^1 Y_2^{-1} +\alpha_2^0 Y_2^0 + \alpha_2^1 Y_2^1 +
    \alpha_2^2 Y_2^2 \right),
\end{aligned} 
\end{equation}
where we expressed $\varepsilon f(\theta,\phi)$ in spherical harmonics. 
For an ellipsoid, the coefficients in front of harmonics $Y_2^1$ and $Y_2^2$ are the same as the coefficients in front of $Y_2^{-1}$ and $Y_2^{-2}$, respectively. The coefficients are
\begin{equation}
    \begin{aligned}
       \alpha_2^0 &= \sqrt{\frac{\pi}{45}}\Big( \epsilon_1\big(1+3\cos(2\beta)\big)+2\epsilon_2 \Big),  \\
       \alpha_2^1 &= -2i\sqrt{\frac{2\pi}{15}} \epsilon_1\cos(\beta)\sin(\beta), \\
       \alpha_2^2 &= -\sqrt{\frac{\pi}{30}} \Big( \epsilon_1 \big(1 - \cos(2\beta) \big) + 2\epsilon_2  \Big).
    \end{aligned}
\label{eq:ell_alphas}
\end{equation}
Note that all $\alpha_2^m=O(\varepsilon)$.

The normal vector and curvature are found by describing the droplet as a level surface $\xi=r - R_0\left(1 + \varepsilon f(\theta,\phi)\right)=0$, then
\begin{equation}
    \vect{n}=\left[\frac{\vect{\nabla} \xi}{|\vect{\nabla} \xi|}\right]_{r=R_0(1+\varepsilon f)},
\end{equation}
and
\begin{equation}
    k_1+k_2=\left[\vect{\nabla} \boldsymbol{\cdot} \vect{n}\right]_{r=R_0(1+\varepsilon f)},
\end{equation}
which we then expand in $\varepsilon$ and $Bm$ keeping only the first order terms. Up to the first order in $\varepsilon$ and $Bm$, $f_M$ remains unchanged from the spherical case.

Just as in the zeroth order solution (for the spherical droplet), the first order correction fields $\varepsilon\ie{\vect{v}}_1, \varepsilon\ie{p}_1,\varepsilon\ar{\vect{v}}_1, \varepsilon\ar{p}_1$ are written in the form of Lamb's solution. The boundary conditions for the full fields ($\vect{v}_0+\varepsilon\vect{v}_1$, $p_0+\varepsilon p_1$) are enforced on the deformed surface $r=R_0(1+\varepsilon f)$ up to the first order of $\varepsilon$ and $Bm$.

Again examining the normal force on the droplet boundary, it becomes evident that the only non-zero coefficients in the Lamb's solution for the first order correction fields are the ones with $l=2$. However, now all $m=-l,...,l$ are needed since there is no longer a symmetry along $\phi$. 
Now there is only a finite number of coefficients, who can be found just as in the spherical case. 
We plug the solution with the non-zero coefficients in the boundary conditions (eq. \eqref{eq:v_BE}, eq. \eqref{eq:normal_force_BE}, eq. \eqref{eq:tang_force_BE}) enforced on the deformed surface $r=R_0(1+\varepsilon f)$, and expand them in power series of $\varepsilon$ and $Bm$ up to first order. 
We then expand the expressions in spherical harmonics and require all the coefficients in front of them to be 0 to satisfy the boundary conditions for all $\theta, \phi$. 

This leads to somewhat long expressions for the velocity and pressure fields inside and outside the droplet (they are provided in the appendix \ref{sec:appendix_deformed}). To get the solution for the case when the magnetic field is rotating, we add $\vect v_{rot}=-\omega\pvec{e}_x \times \vect r$ to the velocity field. 

To describe the change in shape of the droplet, only the velocity on its boundary $\vect{v}^{(b)}=[\vect{v}]_{r=R_0(1+\varepsilon f)}$ is needed.
If the droplet's shape is described by $r=\rho(\theta,\phi,t)=R_0\left(1+\varepsilon f(\theta,\phi,t)\right)$, then the kinematic boundary condition dictates that
\begin{equation}
    \dd{\rho}{t}=v^{(b)}_r-\frac{1}{r}\left( v^{(b)}_\theta \dd{\rho}{\theta} +  \frac{v^{(b)}_\phi}{\sin(\theta)}\dd{\rho}{\phi} \right),
\label{eq:kinematic_BC}
\end{equation}
where again we keep only the terms up to first power of $\varepsilon$ and $Bm$. If we expand the droplet's shape in spherical harmonics $r=\rho(\theta,\phi,t)=R_0\sum_{l=0}^\infty \sum_{m=-l}^l \alpha_l^m (t) Y_l^m(\theta,\phi)$ and plug them in eq. \eqref{eq:kinematic_BC}, we get the differential equations for the expansion coefficients
\begin{equation}
\left\{ \begin{aligned} 
  \frac{d \alpha_2^2}{dt} &= i\omega\alpha_2^1 -\frac{\alpha_2^2}{\tau} \\
  \frac{d \alpha_2^1}{dt} &= \frac{i\omega}{2} \left( \sqrt{6}\alpha_2^0+2\alpha_2^2 \right) - \frac{\alpha_2^1}{\tau}\\
  \frac{d \alpha_2^0}{dt} &=i\sqrt{6}\omega \alpha_2^1+\frac{4\sqrt{5\pi}\delta}{15\tau} - \frac{\alpha_2^0}{\tau}
\end{aligned} \right. .
\label{eq:alpha_diff}
\end{equation}
We also see that the coefficients with opposite signs in $m$ evolve with the same rate $d\alpha_l^m / dt = d\alpha_l^{-m} / dt$. The derivatives of the coefficients with $l\neq2$ are zero. Which means that an initially ellipsoidal droplet stays ellipsoidal. 

If we insert the ellipsoid expansion coefficients from eq. \eqref{eq:ell_alphas} into eq. \eqref{eq:alpha_diff}, we retrieve the system of equations for the time evolution of the ellipsoid deformation parameters (eq. \eqref{eq:eps_system}). Furthermore, from the hydrodynamic approach we also get the expressions for the phenomenological constants in anisotropy tensor approach. The elongation of the droplet in a non-rotating field is
\begin{equation}
\begin{split}
    \delta
    =& \frac{9 Bm}{32\pi} \frac{\left(\mu_r-1\right)^2}{\left(\mu_r+2\right)^2},
\label{eq:delta_expression}
\end{split}
\end{equation}
and the characteristic relaxation time for small perturbations from equilibrium is
\begin{equation}
    \tau=\frac{R_0\ar{\eta}}{\gamma}\frac{(3+2\lambda)(16+19\lambda)}{40(1+\lambda)}.
\label{eq:tau_expression}
\end{equation}
The values obtained here for $\delta$ agree in the limit of small deformations with the equilibrium values of an ellipsoidal droplet in a homogeneous static field \cite{bacri-instability-1982, afkhami-deformation-2010, tsebers-virial-1985}. And $\tau$ is the same as described in other works on small deformations of droplets \cite{dikanskii-magnetic-1990, taylor-formation-1934}.

\section{Analysis and comparison with numerical simulations}
\label{sec:analysis}
\subsection{Fixed points and their stability}
The fixed points for eq. \eqref{eq:eps_system} are
\begin{equation}
\left\{ \begin{aligned} 
  \epsilon_1^*&= \frac{\delta}{\sqrt{1+4\tau^2\omega^2}} \\
  \epsilon_2^*&= \frac{\delta}{2}\left(1 - \frac{1}{\sqrt{1+4\tau^2\omega^2}}  \right)\\
  \beta^* &= -\frac{1}{2}\arctan(2 \tau \omega )+n\pi
\end{aligned} \right. ,
\label{eq:fixed_pts}
\end{equation}
where $n$ is a whole number. 
We see that for non-rotating field $\tau\omega=0$, the droplet becomes prolate $\epsilon_2^*=0$, and aligns with the field $\beta^*=n\pi$. 
Whereas for large field rotation frequencies $\tau\omega \rightarrow \infty$, the droplet becomes nearly oblate $\epsilon_1^* \rightarrow 0$, it flattens to a value of $\epsilon_2^* \rightarrow \delta / 2$ and lags the field by an angle $\beta^*\rightarrow -\pi / 4 + n\pi$.

It is interesting to investigate the stability of the fixed point. Linearizing the equations \eqref{eq:eps_system} for small perturbations $\Delta\epsilon_1$, $\Delta\epsilon_2$, $\Delta\beta$ around the fixed point, we get
\begin{equation}
    \frac{d}{dt}
    \begin{pmatrix}
    \Delta\epsilon_1\\
    \Delta\epsilon_2\\
    \Delta\beta
    \end{pmatrix}
    =
    \begin{pmatrix}
    -\frac{1}{\tau} & 0 & \frac{4\delta\omega}{\sqrt{1+4\tau^2\omega^2}} \\
    0 & -\frac{1}{\tau} & -\frac{2\delta\omega}{\sqrt{1+4\tau^2\omega^2}} \\
    -\frac{\omega\sqrt{1+4\tau^2\omega^2}}{\delta} & 0 & -\frac{1}{\tau}
    
    \end{pmatrix}
    \begin{pmatrix}
    \Delta\epsilon_1\\
    \Delta\epsilon_2\\
    \Delta\beta
    \end{pmatrix}.
\end{equation}
The eigenvalues of this matrix are $-1 / \tau$, $-1 / \tau -2i\omega$, $-1 / \tau +2i\omega$, meaning that the fixed points are stable foci, and there is an oscillation with twice the angular frequency of the magnetic field as the perturbations decay.

\subsection{Qualitative behavior of the system}

We can scale the deformation parameters by the deformation parameter in a non-rotating field $\delta$ (eq. \eqref{eq:delta_expression}) and time by the characteristic decay time $\tau$ (eq. \eqref{eq:tau_expression}), setting $\tilde\epsilon_1 = \epsilon_1/\delta$, $\tilde\epsilon_2 = \epsilon_2/\delta$ and  $\tilde t = t/\tau$. We then see that up to a constant scaling factor the system is governed only by a single free parameter $\tau\omega$, which is proportional to the capillary number - the ratio of viscous forces to the surface tension forces \cite{stone-dynamics-1994}. In this form, the system \eqref{eq:eps_system} is written as

\begin{equation}
\left\{ \begin{aligned} 
  \frac{d \tilde\epsilon_1}{d \tilde t} &= - \tilde\epsilon_1 +  \cos(2\beta)  \\
  \frac{d \tilde\epsilon_2}{d \tilde t} &= - \tilde\epsilon_2 +  \sin^2(\beta) \\
  \frac{d \beta}{d \tilde t} &= -\tau\omega - \frac{ \cos(\beta)\sin(\beta)}{ \tilde\epsilon_1}
\end{aligned} \right. .
\label{eq:eps_scaled_system}
\end{equation}

\begin{figure}[h]
    \centering
    \begin{overpic}[width=0.45\textwidth]{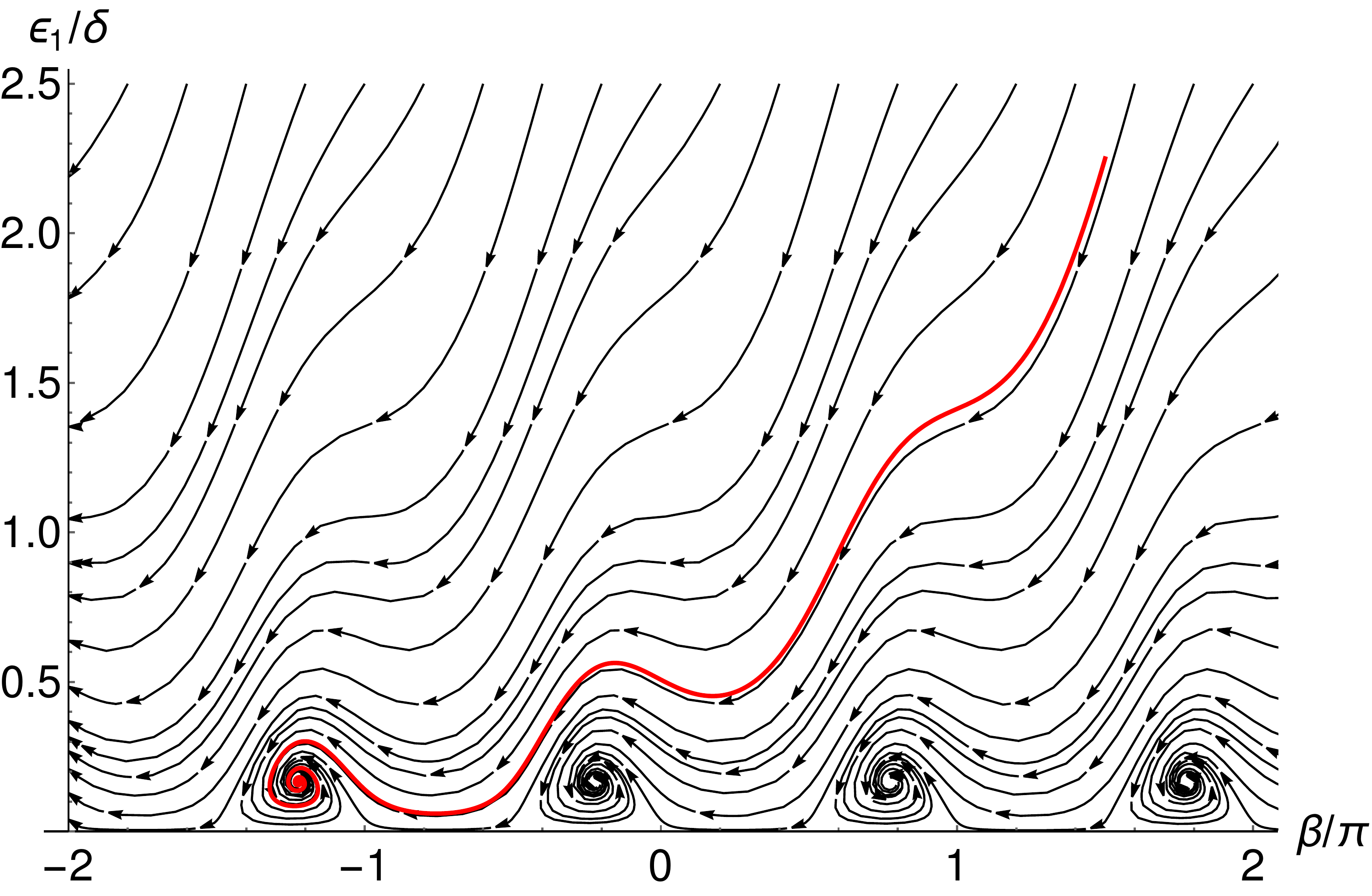}
        \put (-5,62) {\textcolor{black}{(a)}}
    \end{overpic}
    \begin{overpic}[width=0.45\textwidth]{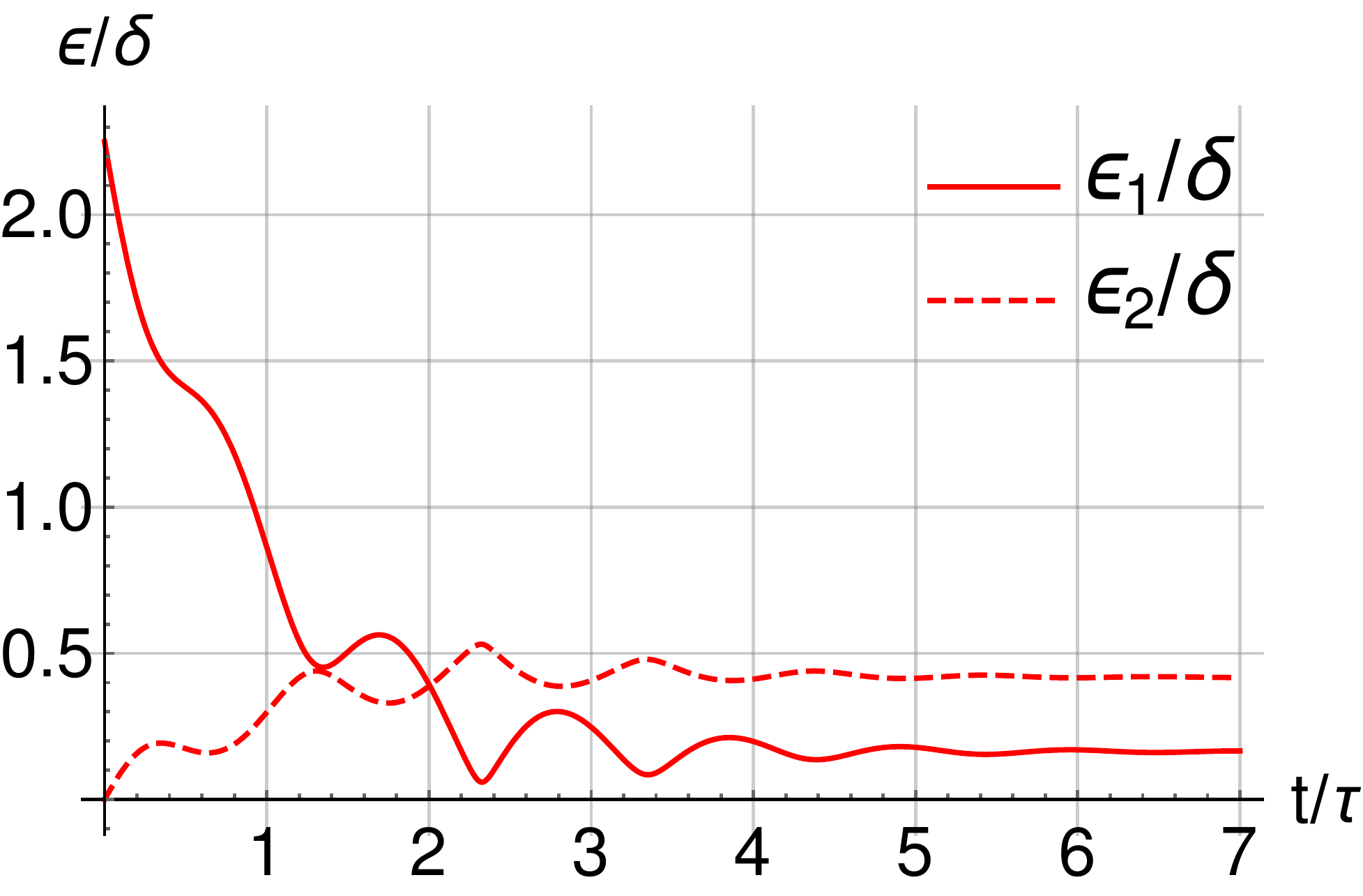}
        \put (-5,60) {\textcolor{black}{(b)}}
    \end{overpic}
    \begin{overpic}[width=0.45\textwidth]{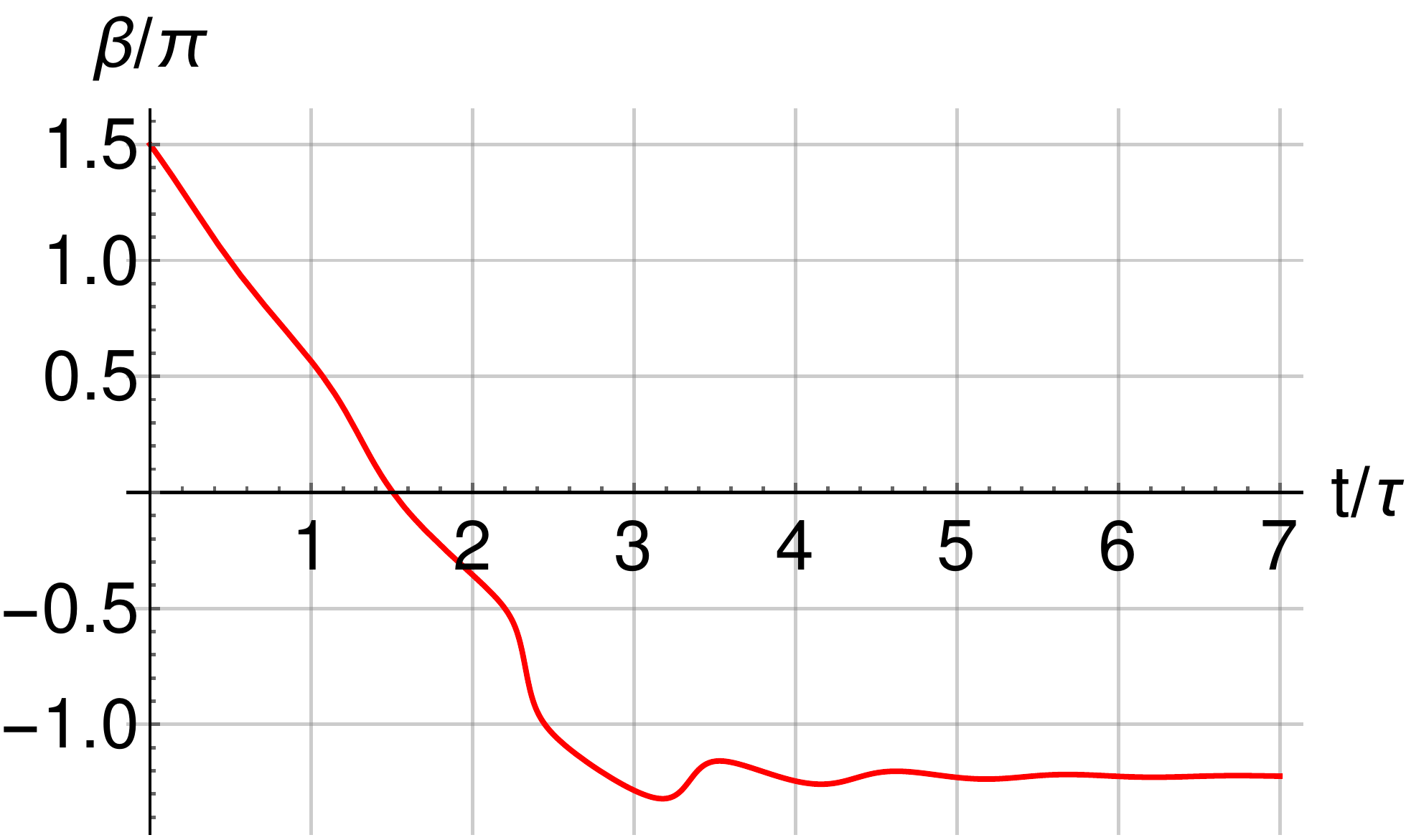}
        \put (-5,52) {\textcolor{black}{(c)}}
    \end{overpic}
    \caption{(a) shows the phase portrait of the parameter describing the droplet elongation relative to its deformation under non-rotating field $\epsilon_1/\delta$ and the angle between the droplet and the magnetic field $\beta$. The field frequency is such that $\tau\omega=3$. (b) and (c) show the time evolution of the variables $\epsilon_1/\delta$, $\epsilon_2/\delta$ and $\beta$ corresponding to the red trajectory in (a).
    }
    \label{fig:phase_portrait}
\end{figure}

The variables $\epsilon_1$ and $\beta$ form a closed system. We can use that to visualize their evolution by drawing a phase portrait (figure \ref{fig:phase_portrait}). Similarities can be seen with the behavior of a damped pendulum, where if the initial elongation is large enough, the droplet overshoots its equilibrium angle $\beta$ several times before settling. 
Unlike the damped pendulum, in this system there is no inertia - it is completely dissipative. However, similarly to the pendulum, there is an interplay between two characteristic times - the oscillatory period $T=\pi / \omega$ (the shape oscillates with twice the frequency of the magnetic field) and the small deformation decay time $\tau$.

\begin{figure*}[ht]
    \centering
    \includegraphics[width=0.9\textwidth]{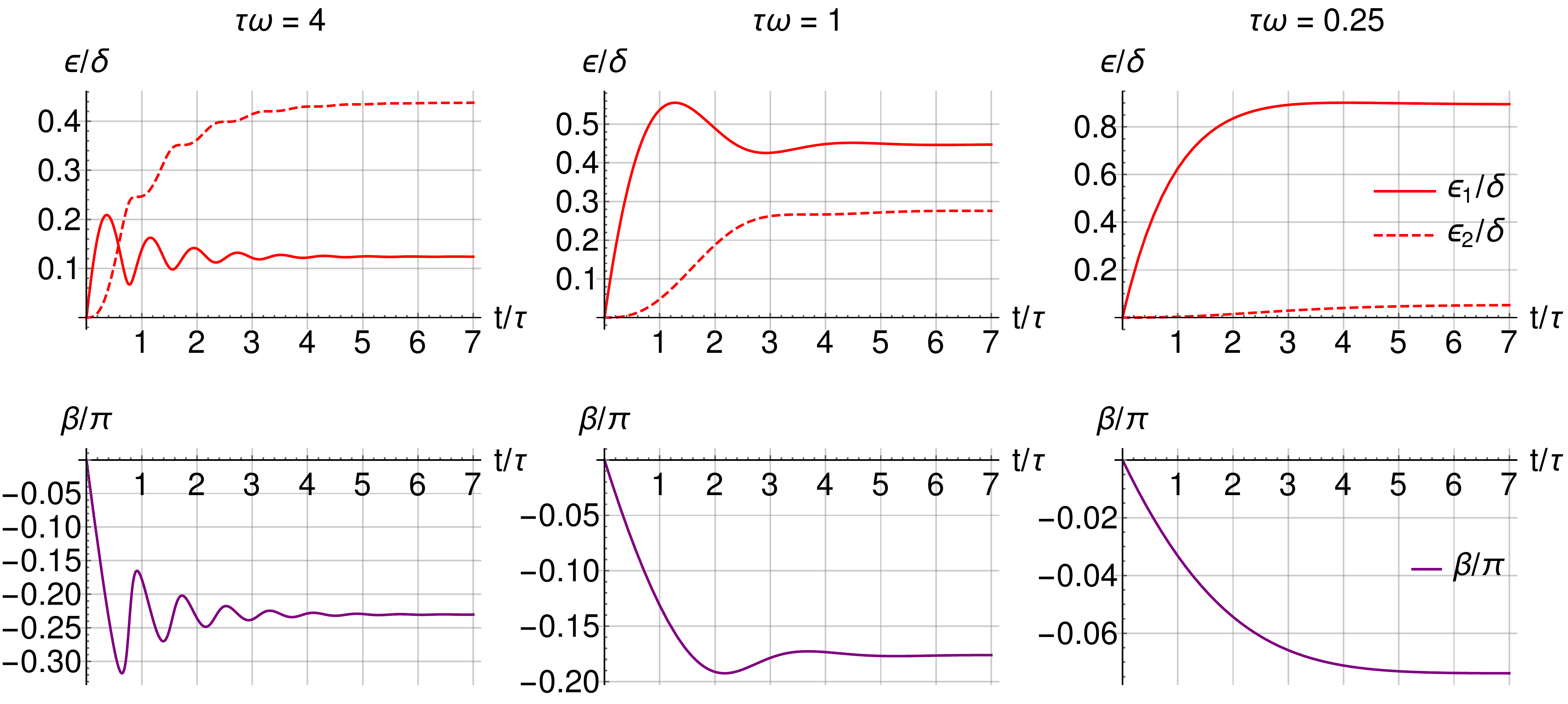}
    \caption{Relaxation to an equilibrium shape of an initially spherical droplet when a rotating magnetic field of different frequencies $\tau\omega$ is applied. The top row shows the deformation parameters $\epsilon_1$ and $\epsilon_2$ and the bottom row shows the angle $\beta$ between the droplet's largest axis and the field.
    }
    \label{fig:tauomega}
\end{figure*}

The transient behavior of an initially spherical droplet when placed in a rotating magnetic field illustrates the interplay between the two characteristic times of the system (figure \ref{fig:tauomega}). 
If the oscillatory period $T$ is small compared to the relaxation time $\tau$ ($\tau\omega \gg 1$), there is enough time for droplet to make several oscillations before settling, however, if the the opposite is the case, the droplet relaxes to the equilibrium before a single oscillation can occur.

\subsection{Flow fields around the droplet}

\begin{figure*}[ht]
    \begin{subfigure}[c]{.32\textwidth}
    \begin{overpic}[width=\textwidth]{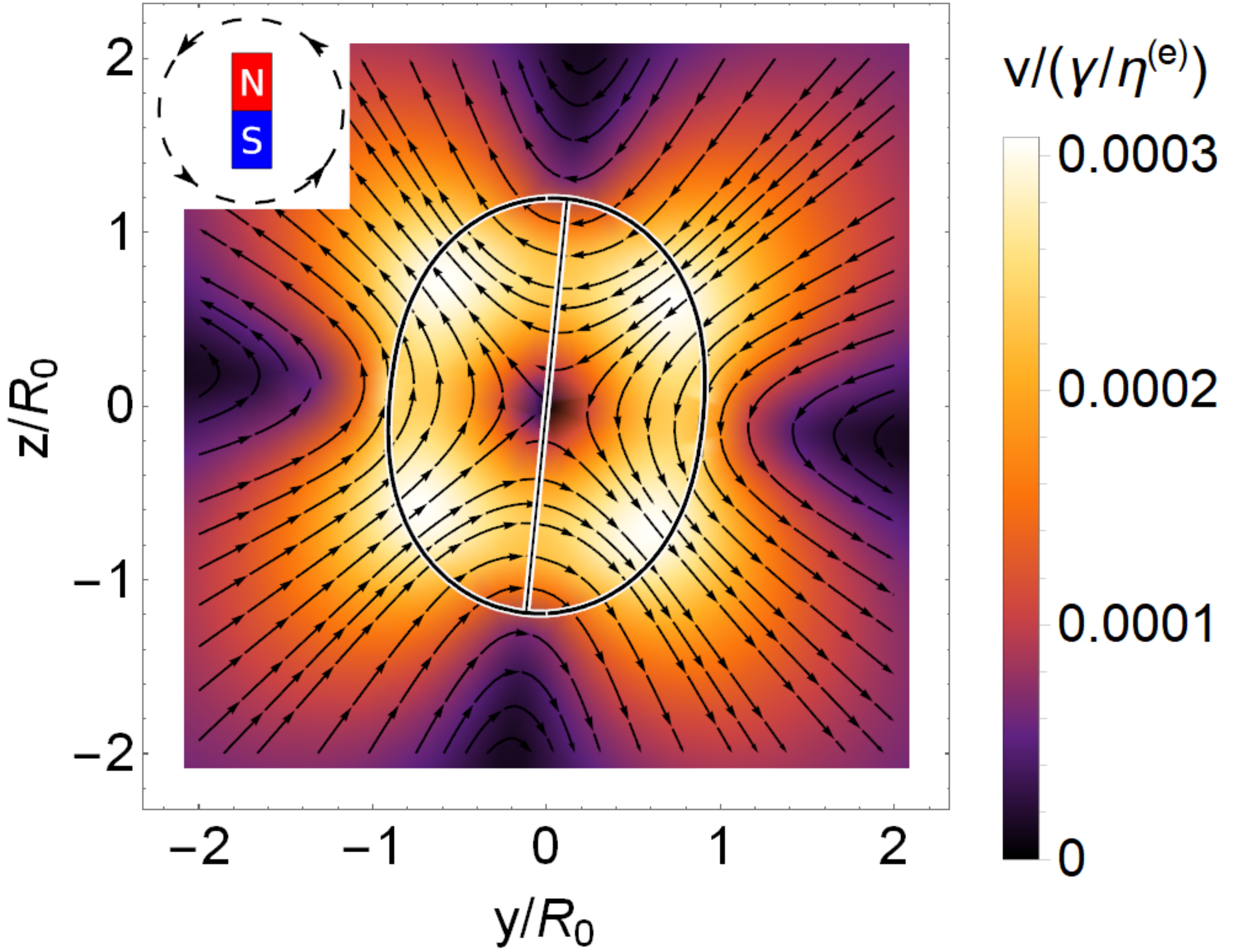}
        \put (-1,70) {\textcolor{black}{(a)}}
    \end{overpic}
    \end{subfigure}
    \begin{subfigure}[c]{.32\textwidth}
    \begin{overpic}[width=\textwidth]{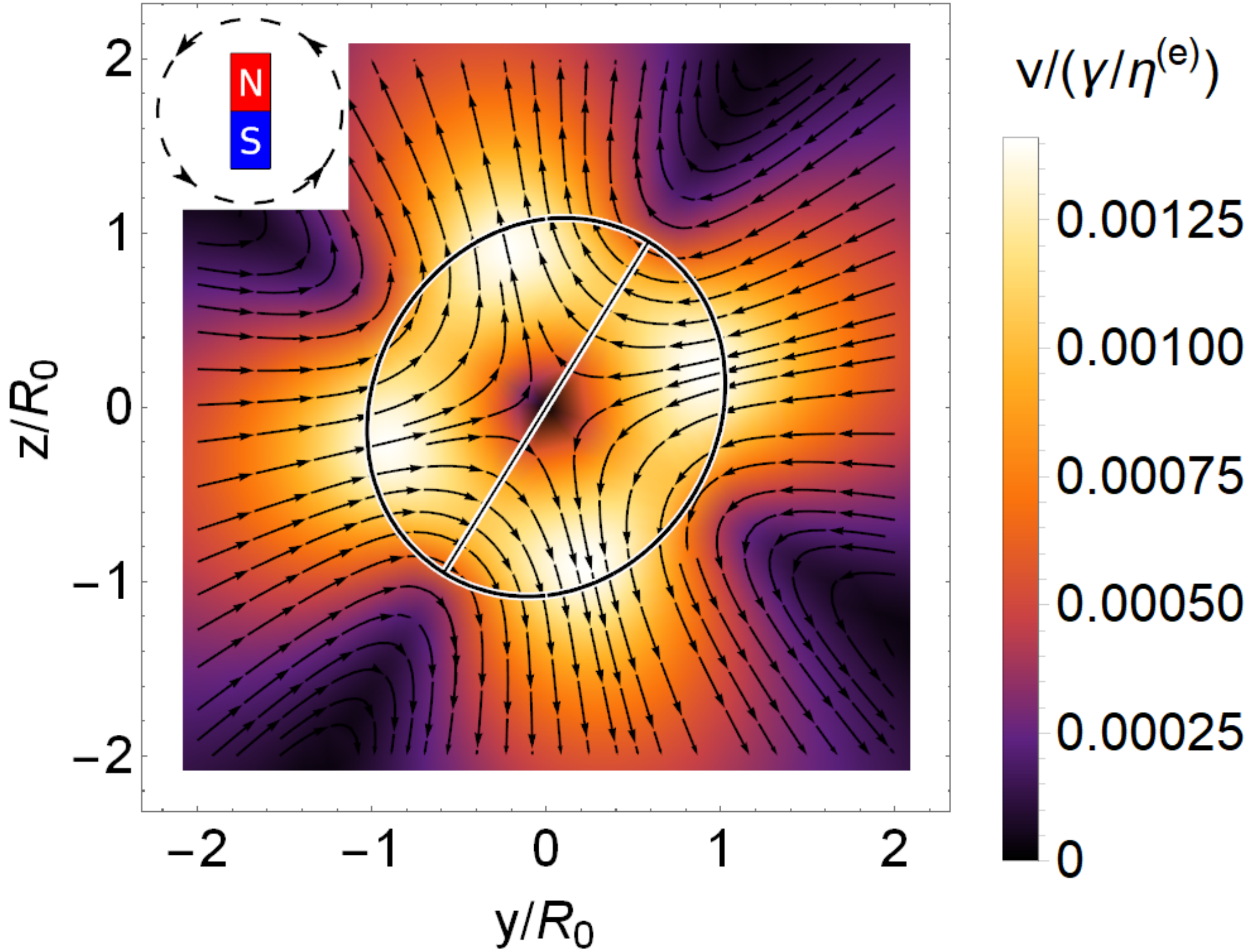}
        \put (-1,70) {\textcolor{black}{(b)}}
    \end{overpic}
    %
    \end{subfigure}
    \begin{subfigure}[c]{.32\textwidth}
    \begin{overpic}[width=\textwidth]{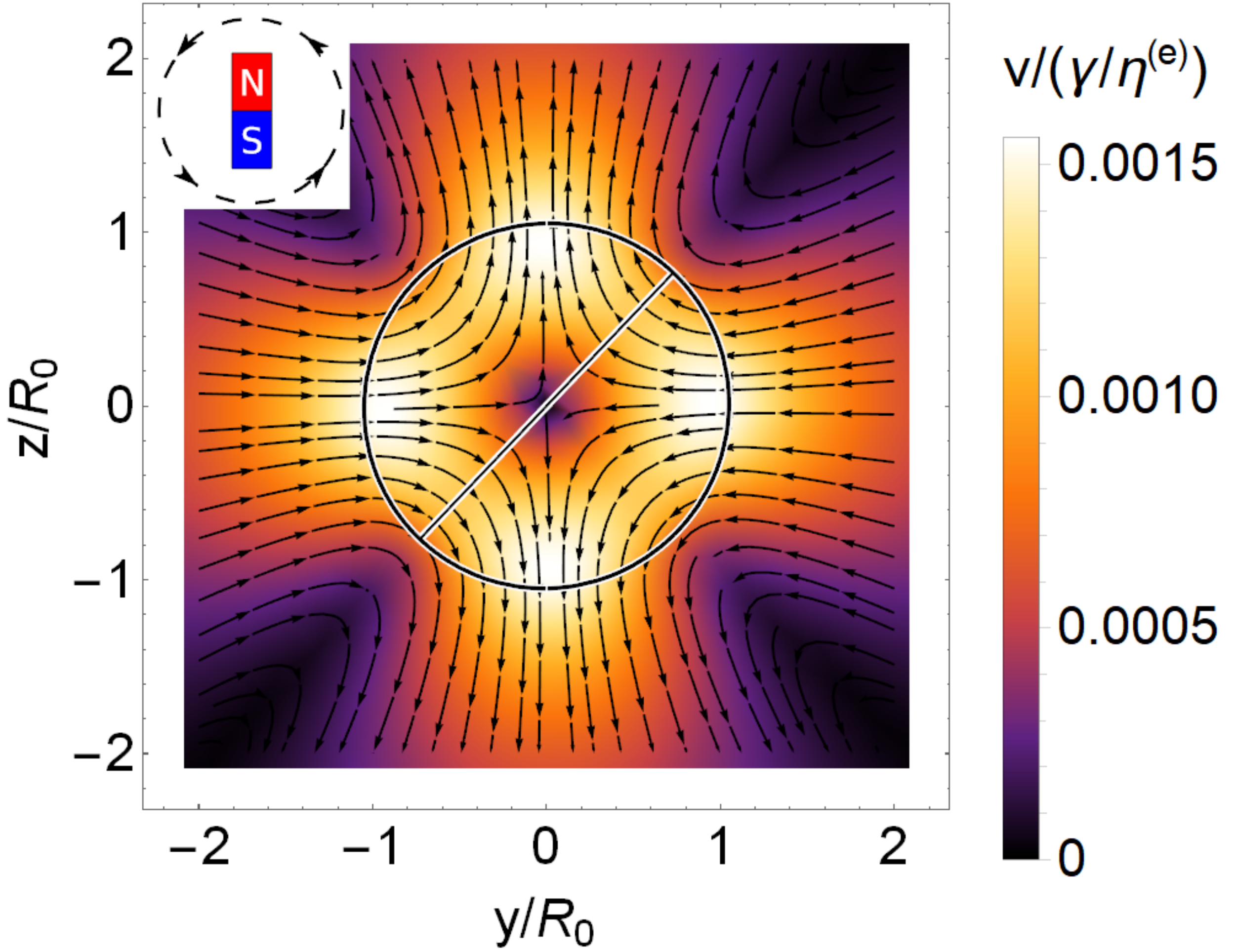}
        \put (-1,70) {\textcolor{black}{(c)}}
    \end{overpic}
    \end{subfigure}
    
    \caption{The velocity field cross-section at $x=0$ inside and outside the magnetic droplets in the laboratory reference frame. The magnetic field is momentarily pointing in the $z$ axis direction and rotating counterclockwise. The plots are made with equilibrium shapes of the droplets with $\delta=0.3$ and (a) $\tau\omega=0.1$, (b) $\tau\omega=1$ and (c) $\tau\omega=10$. The droplet shapes and their longest axis are outlined. The flow is such that it turns the longest axis of the droplet counterclockwise. The velocity magnitude in units of $\gamma/\ar{\eta}$ is shown with a color gradient.}
    \label{fig:flow_field}
\end{figure*}

The solution using the hydrodynamic approach allows us to visualize the flow fields inside and outside the droplets. 
Figure \ref{fig:flow_field} shows the velocity streamlines in the stationary frame of reference for different droplet equilibrium shapes. 

A visual inspection of the flow fields shows that the droplets are not in a rigid body rotation, but rather are changing their orientation due to surface deformations.
Indeed if we integrate the outer velocity and pressure field over the droplet surface, we get that the net torque the droplet exerts on the fluid is identically zero up to first order in $\varepsilon$. 

The magnetic torque exerted by the droplet is $\mu_0 (4\pi R_0^3/ 3) \vect{M}\times \vect H_\infty $, which is proportional to the deformation of the droplet ($\epsilon_1$, $\epsilon_2$) multiplied by $Bm$.
From the equation \eqref{eq:fixed_pts} it can be seen that the deformation of the droplet  is proportional to $Bm$. 
But it then follows that the magnetic torque is proportional to $Bm^2$.
To consider a real rotation of the droplet that imparts a torque, we must go beyond the first order terms in the solution.

A similar result was found in a work that experimentally and with simulations examined sessile water droplets in a rotating electric field \cite{ghazian-oscillation-2014}.
It was found that just as here, the droplets appear to rotate with the angular velocity of the rotating field, but the internal flow fields produce this apparent rotation by deforming the droplet's surface.
The authors of said work called this motion ``pseudo-rotation".
This result can be contrasted with the Quincke rotation (or electrorotation) of weakly conducting droplets where rotational rotlet-like velocity fields emerge as the rotation starts \cite{das-three-dimensional-2021,vlahovska-electrohydrodynamics-2019}.

\subsection{Comparison with numerical calculations}
We use a numerical algorithm based on the boundary element method (BEM) that calculates the 3D evolution of a magnetic droplet's shape in an arbitrary magnetic field. The algorithm (outlined in the appendix \ref{sec:BEM}) is an extension of the work by Erdmanis et al.\cite{erdmanis-magnetic-2017} to be able to capture the dynamics of droplets with $\lambda\neq1$. 

\begin{figure}[ht]
    \centering
    \includegraphics[width=0.4\textwidth]{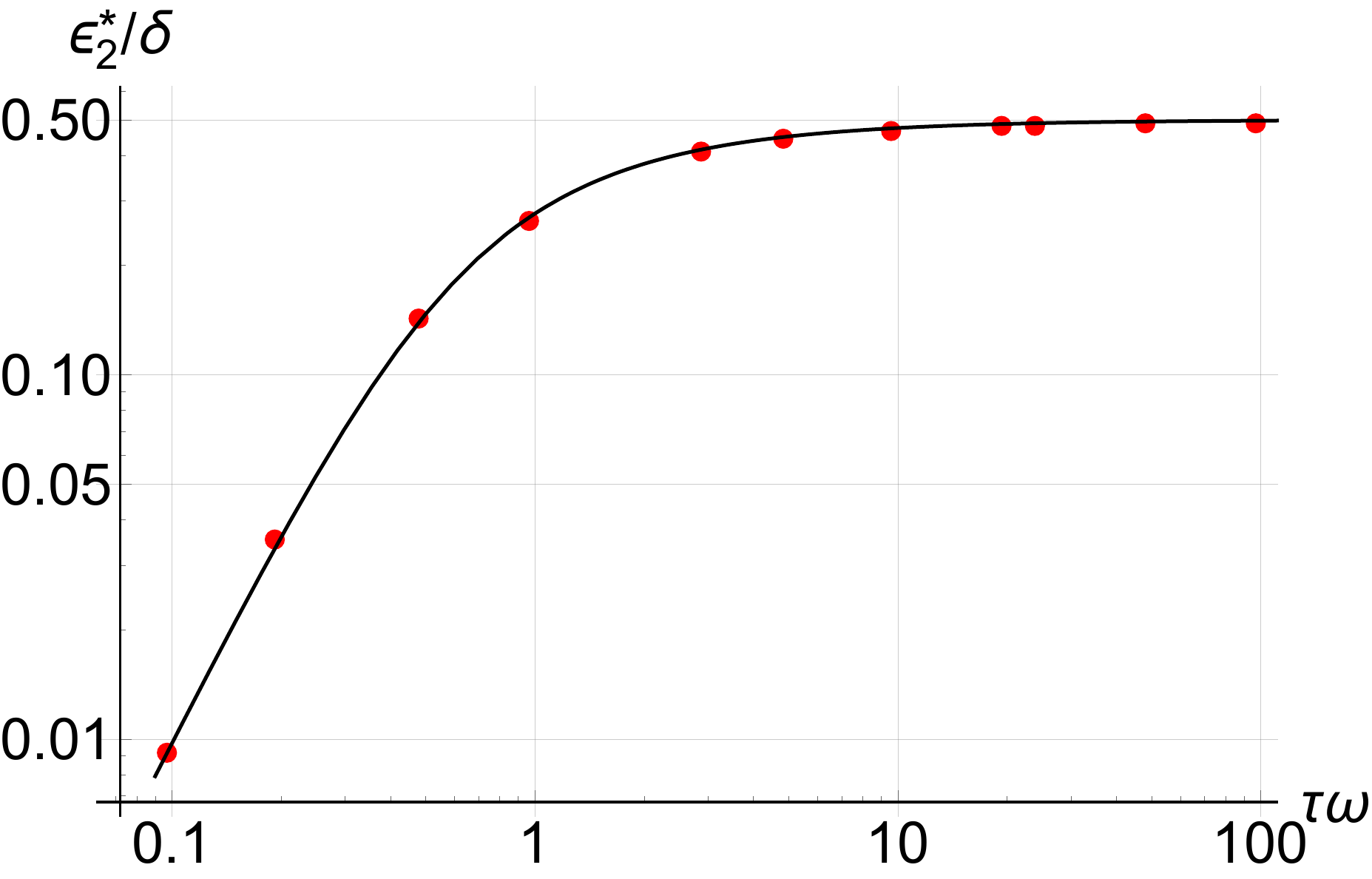}
    \includegraphics[width=0.4\textwidth]{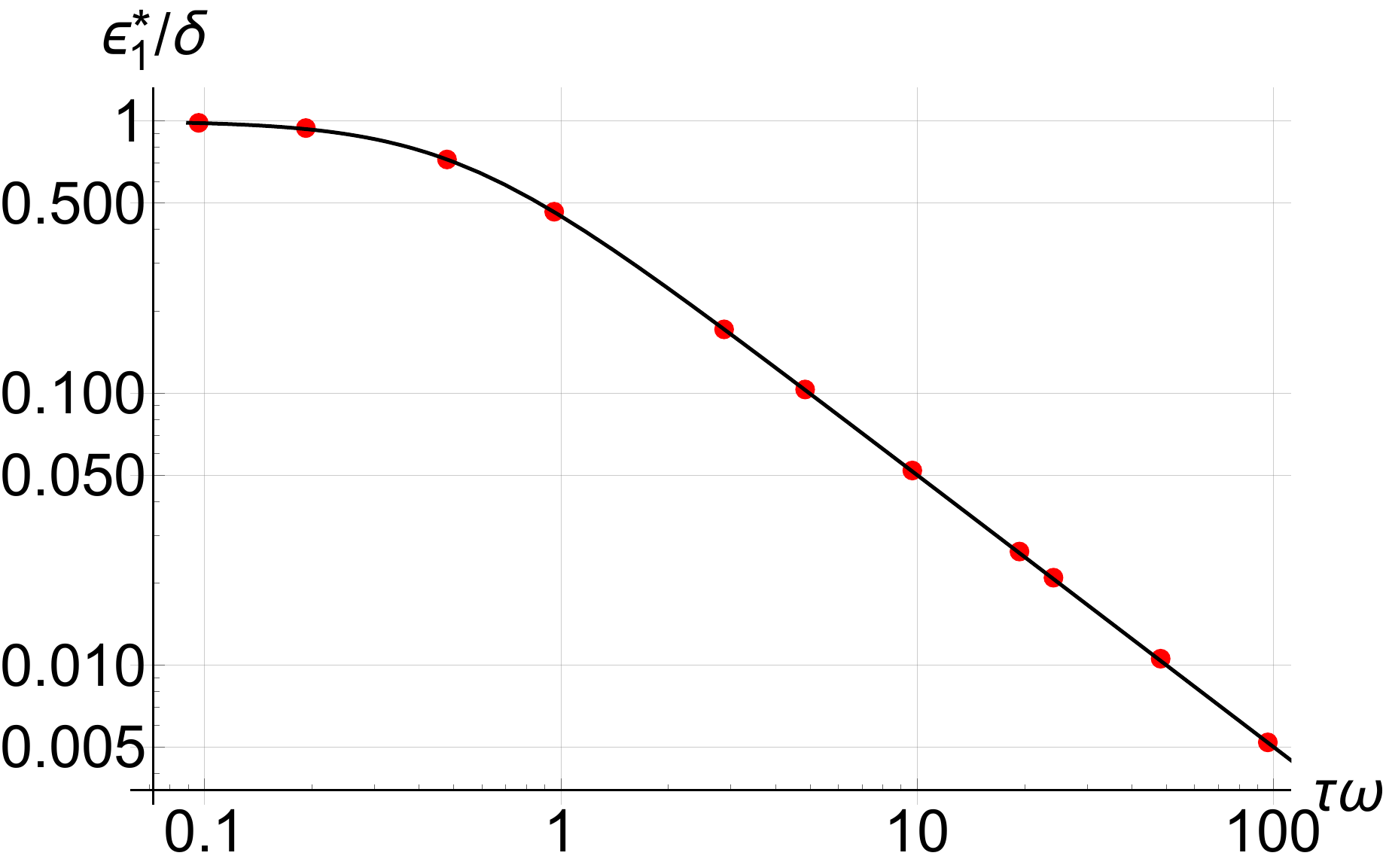}
    \includegraphics[width=0.4\textwidth]{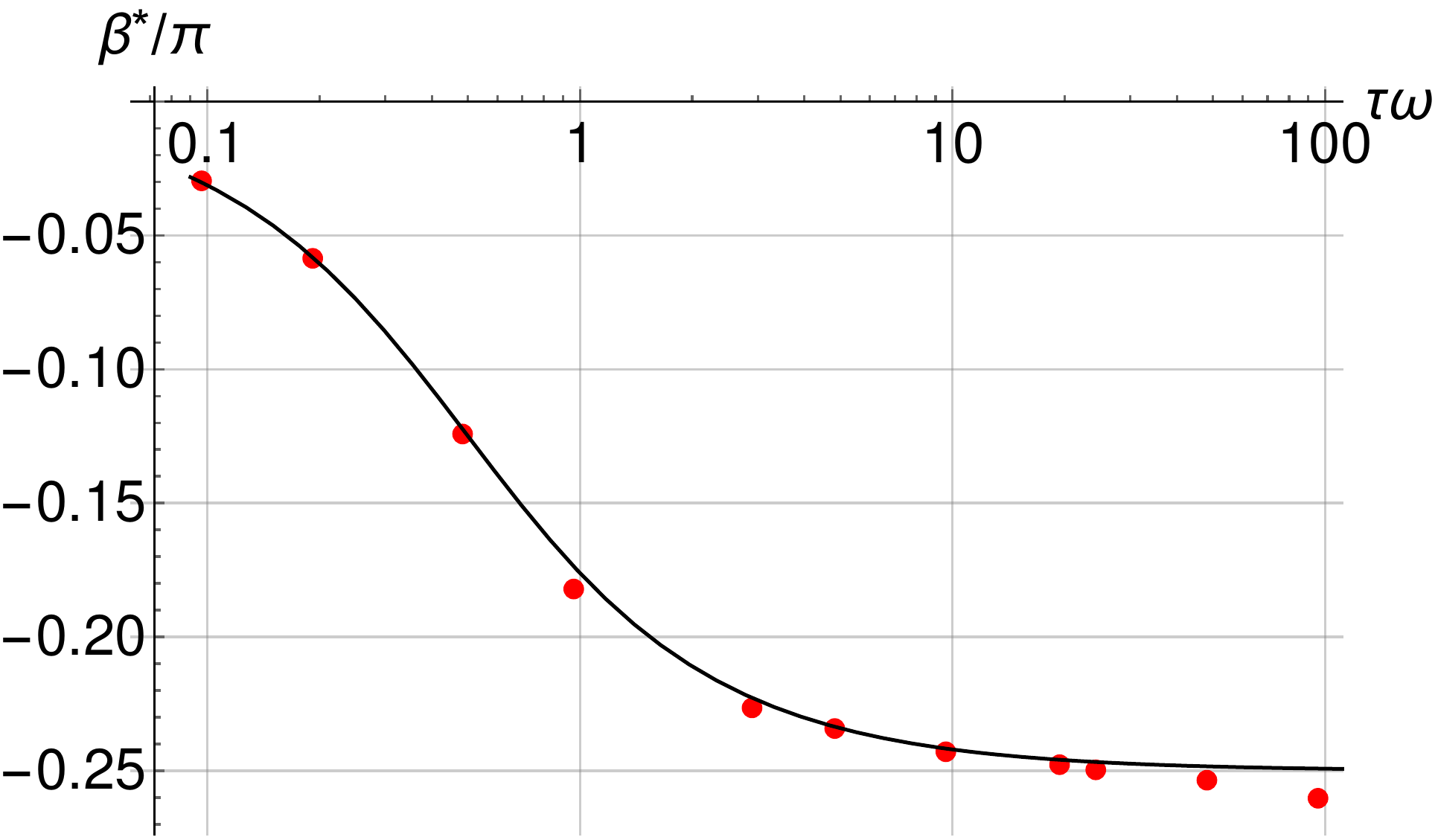}
    \caption{The equilibrium values of the elongation $\epsilon_1^*$, flatness $\epsilon_2^*$ and the angle between the field and the droplet's largest axis $\beta^*$ depending on the field rotation frequency. The droplet elongation in non-rotating field is $\delta=0.005$. The red dots are BEM simulation results and the black lines are from \eqref{eq:fixed_pts}.}
    \label{fig:numerics_vs_analytics}
\end{figure}

To validate the results of the small deformation theory, we calculate the equilibrium shapes using the BEM algorithm and compare them with the analytic expressions (eq. \eqref{eq:fixed_pts}) for the fixed points (figure \ref{fig:numerics_vs_analytics}). 
The dimensionless input parameters for the simulation are chosen as follows: $Bm=0.1$, $\lambda=100$, $\mu_r=10$ and $\omega/(\gamma/(R_0\ar{\eta}))$ varied. 
These parameters correspond to $\delta=0.005$ and $\tau=96.3/(R_0\ar{\eta}/\gamma)$. 
To get the deformation parameters and the angle between the field and the droplet's largest axis, we fit a 3D ellipsoid to the vertices of the mesh triangles. 
We see excellent agreement for both deformation parameters $\epsilon^*_1$ and $\epsilon^*_2$, and a good agreement for $\beta^*$. 
Possibly, the discrepancy between the theoretical curve of $\beta$ and simulation results for large $\omega$ are due to elongation $\epsilon_1$ tending to 0 and thus the angle between the largest axis and the field $\beta$ becomes ill-defined.

\begin{figure}[h]
    \centering
    \includegraphics[width=0.4\textwidth]{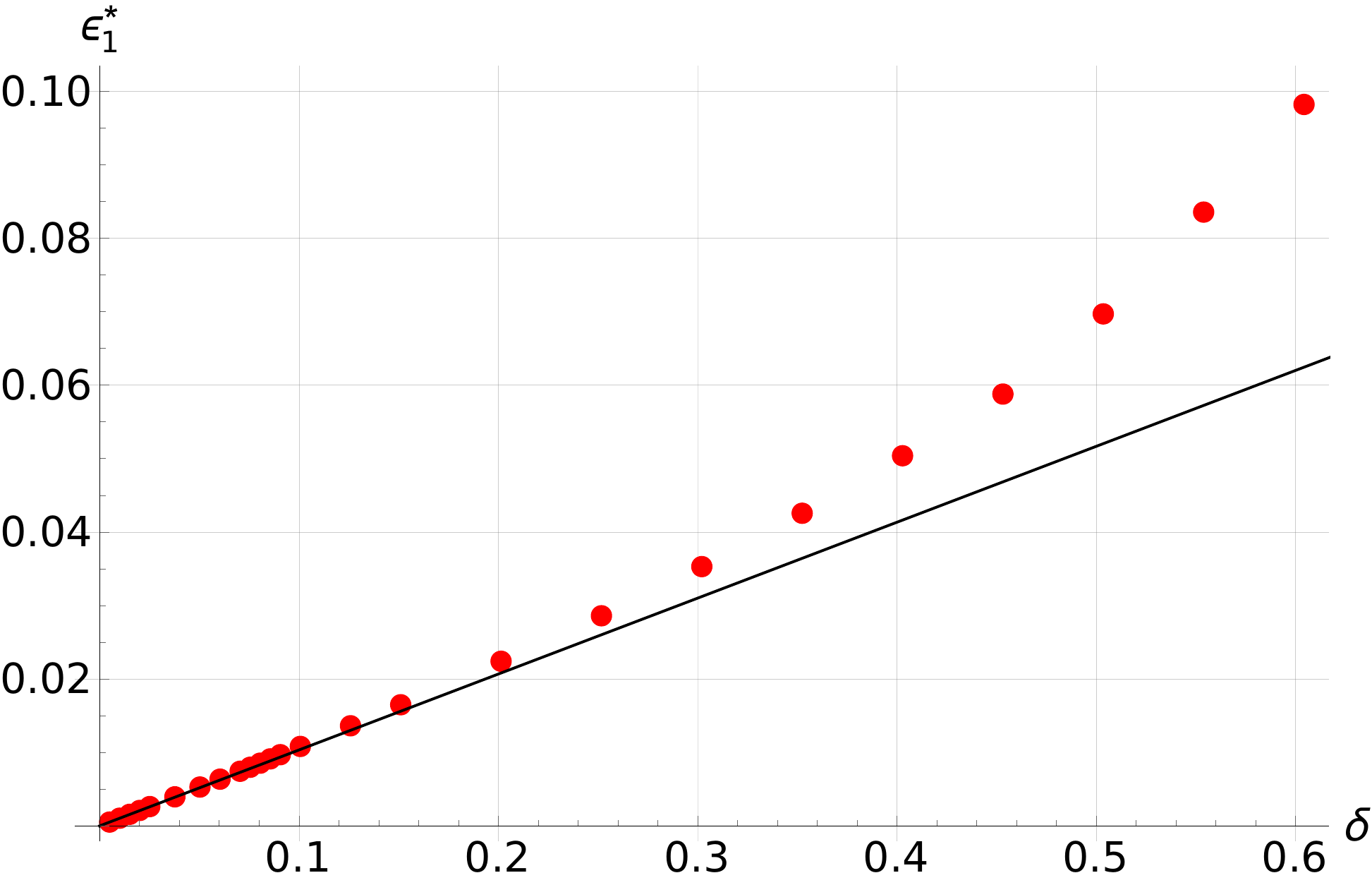}
    \includegraphics[width=0.4\textwidth]{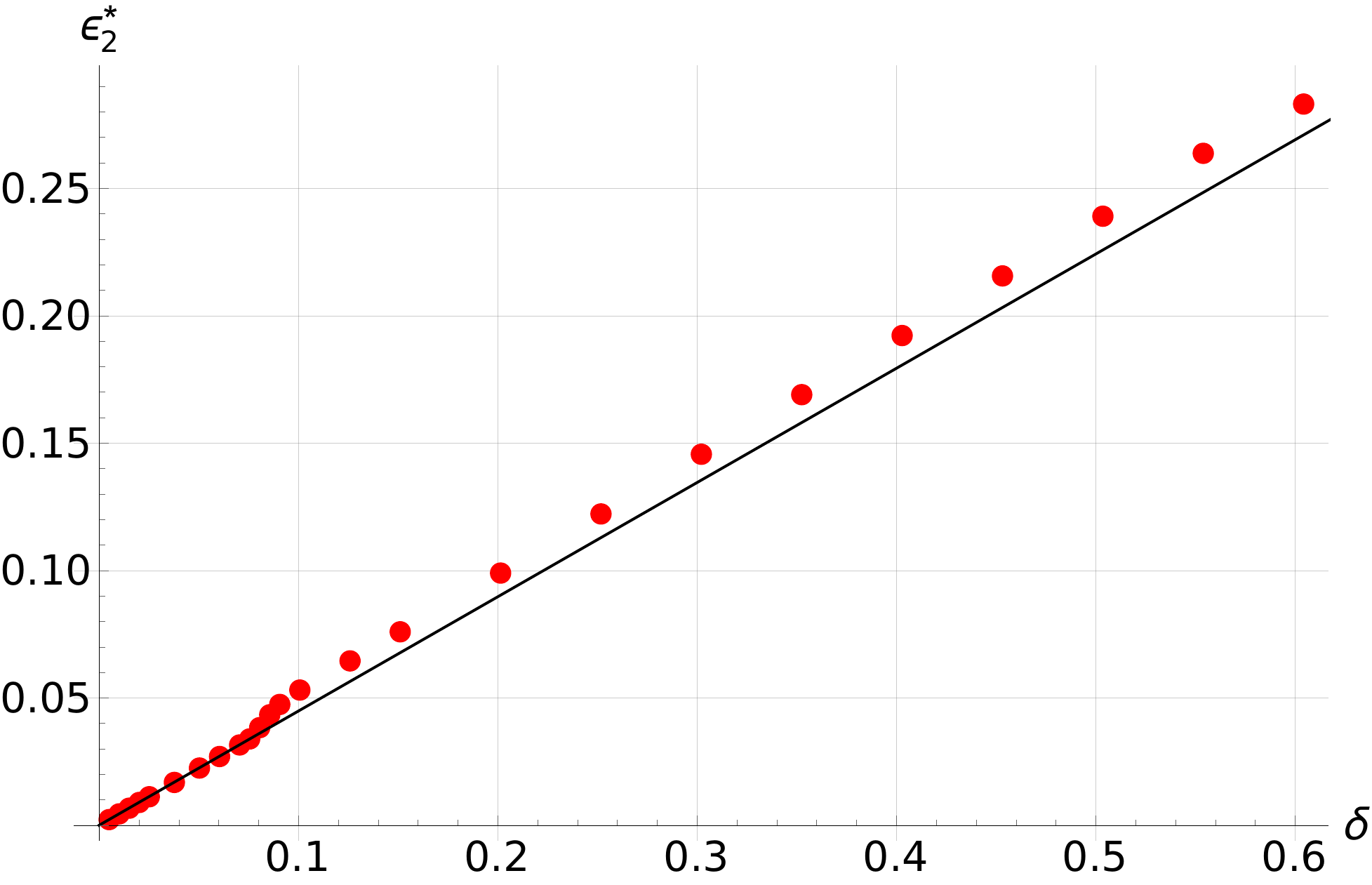}
    \caption{The equilibrium values of the elongation $\epsilon_1^*$, flatness $\epsilon_2^*$ depending on the scaled magnetic field characterized by the parameter $\delta$ as calculated from eq. \eqref{eq:delta_expression}. The droplet's characteristic deformation time multiplied by the field rotation frequency is $\tau\omega=4.8$. The red dots are BEM simulation results and the black lines are from eq. \eqref{eq:fixed_pts}.}
    \label{fig:theory_lim}
\end{figure}

To determine the limits of the small deformation theory, we numerically calculate the equilibrium shapes for increasing values of $Bm$ (proportional to the parameter $\delta$ as given by eq. \eqref{eq:delta_expression}) and compare them to eq. \eqref{eq:fixed_pts} (figure \ref{fig:theory_lim}). 
The simulation parameters are $\omega=0.05/(\gamma/(R_0\ar{\eta}))$, $\lambda=100$, $\mu_r=10$ and $Bm$ is varied from 0.1 to 13.
These parameters correspond to $\tau=96.3/(R_0\ar{\eta}/\gamma)$. 
We see that the error is roughly below $10\%$, if $\delta<0.3$.
Indeed we see that although $Bm\propto\delta$, instead of $Bm\ll 1$, the criterion which should be assessed to determine if the small deformation theory is applicable for a particular case is $\delta\ll 1$.

In this comparison between simulations and the theory, the droplet is quasi oblate $\epsilon_2>\epsilon_1$. 
There is a qualitative agreement in droplet's behavior between the theory and simulations also for values of $Bm\lesssim 12$ ($\delta\lesssim 0.6$).
However, for these simulation parameters, at around $Bm\approx13$ ($\delta\approx 0.65$) the droplet ceases to be quasi oblate and strongly elongates ($\epsilon_1\approx2.7, \epsilon_2\approx0.03$) in roughly the field direction and rotates more or less like a rigid body. 
Such a bifurcation is not captured by the $O(\varepsilon)$ small deformation theory but it has been observed experimentally (left column of figure 1 in the work by Bacri et al.\cite{bacri-behavior-1994}).

\section{Conclusions}
\label{sec:conclusions}

We have produced an analytic 3D solution for the magnetic droplet shape dynamics in a rotating field valid for small deformations and small $Bm$ in the leading order. 
In particular, the parameter $\delta \lesssim 0.3$ as expressed in eq. \eqref{eq:delta_expression} is what determines the limit of this small deformation theory.
When the droplet is elongated in the plane of the rotating field, its shape evolution is governed by a system of three nonlinear differential equations (eq. \eqref{eq:eps_system}).
Its solution is determined up to a scaling by a single parameter $\tau\omega$ - the product of the decay time of small deformations and the magnetic field rotation frequency.

The hydrodynamic equations governing the droplet are completely dissipative, nonetheless a droplet in a rotating field can experience nonlinear damped oscillations before reaching an equilibrium shape.
The interplay of two characteristic times (the field rotation period and the droplet relaxation time) leads to a phase portrait similar to that of a damped pendulum.
Interestingly, for weak fields the droplets pseudo-rotate in the direction of the magnetic field - their surface deforms such that
the long axis follows the field, nonetheless no torque is exerted by the droplet.

We have showcased, how the anisotropy tensor description can be used to calculate the droplet shape in rotating (or precessing) magnetic field.
The phenomenological model was validated with numerical simulations and by solving the full hydrodynamic problem.
In the limit of small deformations the phenomenological equation of motion (eq. \eqref{eq:zeta_eq}) is exact.

Our results could be used for the verification of numerical algorithms and could serve as a basis for more complex models, for example, ones that incorporate a shear flow \cite{maffettone_equation_1998}, which could be then used to analytically calculate the rheological properties of magnetic droplets, which has so far been tackled numerically \cite{cunha-effects-2020,ishida-rheology-2020}.

\begin{acknowledgments}
A.P.S. is thankful to SIA ``Mikrotīkls" and the Embassy of France in Latvia for financially supporting cotutelle studies.

A.P.S. acknowledges the financial support of ``Strengthening of the capacity of doctoral studies at the University of Latvia within the framework of the new doctoral model", identification No. 8.2.2.0/20/I/006

A.C. and A.P.S. acknowledge the financial support of grant of Scientific Council of Latvia lzp-2020/1-0149.
\end{acknowledgments}

\section*{Author declarations}
\subsection*{Conflict of interest}
The authors have no conflicts of interest to declare.
\section*{Data availability}
The data that support the findings of this study are available from the corresponding author upon reasonable request.



%
%

%


\appendix

\section{Velocity and effective pressure of a spherical droplet}
\label{sec:appendix_spherical}
In spherical coordinates
\begin{equation}
\left\{ \begin{aligned} 
    x &= r \cos(\phi)\sin(\theta) \\
    y &= r \sin(\phi)\sin(\theta) \\
    z &= r\cos(\phi)
\end{aligned} \right. ,
\end{equation}
if the magnetic field is pointing in the $\pvec{e}_z$ axis direction, the velocity and effective pressure inside $(i)$ and outside $(e)$ a spherical magnetic droplet read

\begin{equation}
\begin{split}
    &\ie{\vect{v}}_0 = \frac{\gamma}{\ar{\eta}} M_2\Bigg[\\
    &  \left( \frac{L_4}{L_2L_3}\left(\frac{r}{R_0}\right) -\frac{3}{L_3} \left(\frac{r}{R_0}\right)^3 \right)\left( 1+3\cos(2\theta)\right)\vect{e}_r  \\
    &+ \left(-\frac{3L_4}{L_2L_3} \left(\frac{r}{R_0}\right) +\frac{15}{L_3} \left(\frac{r}{R_0}\right)^3\right)\sin(2\theta)\vect{e}_\theta \Bigg],
\end{split}
\end{equation}

\begin{equation}
\begin{split}
    &\ar{\vect{v}}_0 =\frac{\gamma}{\ar{\eta}} M_2\Bigg[\\
    & \left( \frac{1}{L_2} \left(\frac{r}{R_0}\right)^{-2} -\frac{3 L_1}{L_2L_3} \left(\frac{r}{R_0}\right)^{-4} \right)\left( 1+3\cos(2\theta)\right)\vect{e}_r \\
    &+ \left(-\frac{6 L_1}{L_2L_3} \left(\frac{r}{R_0}\right)^{-4} \right)\sin(2\theta)\vect{e}_\theta \Bigg],
\end{split}
\end{equation}

\begin{equation}
    \ie{p}_0 = \frac{\gamma}{R_0}\left( 2 - M_1-\frac{21\lambda M_2}{L_3} \left( 1+3\cos(2\theta)\right) \left(\frac{r}{R_0}\right)^2 \right),
\end{equation}

\begin{equation}
    \ar{p}_0 = \frac{\gamma}{R_0}\frac{2M_2}{L_2}\left( 1+3\cos(2\theta)\right)\left(\frac{r}{R_0}\right)^{-3},
\end{equation}
where 
\begin{equation}
\begin{gathered}
    M_1 = \frac{3Bm}{8\pi}\frac{\mu_r-1}{\mu_r+2}, \quad   M_2 = \frac{3Bm}{16\pi}\frac{\left(\mu_r-1\right)^2}{\left(\mu_r+2\right)^2}, \\
    L_1 = (2+3\lambda), \quad  L_2 = (3+2\lambda), \\
    L_3 = (16+19\lambda), \quad   L_4 = (19+16\lambda). \\
\end{gathered}
\end{equation}

\section{Velocity and effective pressure of a deformed droplet}
\label{sec:appendix_deformed}
The velocity and effective pressure fields in and around a slightly deformed droplet are written as power series
\begin{equation}
\begin{split}
    \ie{\vect{v}}&= \ie{\vect{v}}_0 + \varepsilon\ie{\vect{v}}_1 + O(\varepsilon^2)\\
    \ar{\vect{v}}&= \ar{\vect{v}}_0 + \varepsilon\ar{\vect{v}}_1 + O(\varepsilon^2)\\
    \ie{p}&= \ie{p}_0 + \varepsilon\ie{p}_1 + O(\varepsilon^2)\\
    \ar{p}&= \ar{p}_0 + \varepsilon\ar{p}_1 + O(\varepsilon^2),
\end{split}
\end{equation}
where the first order correction terms are as follows:
\begin{equation}
\begin{split}
    \varepsilon\ie{\vect{v}}_1 =& \frac{\gamma}{\ar{\eta}}\left( -\frac{L_4}{L_2L_3}\left(\frac{r}{R_0}\right) + \frac{3}{L_3} \left(\frac{r}{R_0}\right)^3 \right) E_1 \vect{e}_r \\
    &+\frac{\gamma}{\ar{\eta}}\left( \frac{L_4}{L_2L_3}\left(\frac{r}{R_0}\right) - \frac{5}{L_3} \left(\frac{r}{R_0}\right)^3 \right) E_2 \vect{e}_\theta \\
    &+\frac{\gamma}{\ar{\eta}}\left( \frac{L_4}{L_2L_3}\left(\frac{r}{R_0}\right) - \frac{5}{L_3} \left(\frac{r}{R_0}\right)^3 \right) E_3 \vect{e}_\phi, 
\end{split}
\end{equation}

\begin{equation}
\begin{split}
    \varepsilon\ar{\vect{v}}_1 =& \frac{\gamma}{\ar{\eta}}\left(- \frac{1}{L_2}  \left(\frac{r}{R_0}\right)^{-2} + \frac{3L_1}{L_2L_3}  \left(\frac{r}{R_0}\right)^{-4}  \right) E_1 \vect{e}_r \\
    &+\frac{\gamma}{\ar{\eta}}\frac{2L_1}{L_2L_3} E_2 \vect{e}_\theta \left(\frac{r}{R_0}\right)^{-4} \\
    &+\frac{\gamma}{\ar{\eta}}\frac{2L_1}{L_2L_3} E_3 \vect{e}_\phi \left(\frac{r}{R_0}\right)^{-4}, 
\end{split}
\end{equation}

\begin{equation}
    \varepsilon\ie{p}_1 = \frac{\gamma}{R_0}\frac{21\lambda E_1}{L_3}\left(\frac{r}{R_0}\right)^2,
\end{equation}

\begin{equation}
    \varepsilon\ar{p}_1 = -\frac{\gamma}{R_0}\frac{2E_1}{L_2}\left(\frac{r}{R_0}\right)^{-3},
\end{equation}
where 
\begin{equation}
\begin{split}
    E_1 =& -2 \epsilon_1 \sin (2 \beta ) \sin (2 \theta ) \sin (\phi ) \\
    &+\sin^2(\theta ) \cos (2 \phi ) (\epsilon_1 \cos (2 \beta
   )-\epsilon_1-2 \epsilon_2) \\
   &+\frac{1}{6} (3 \cos (2 \theta
   )+1) (3 \epsilon_1 \cos (2 \beta )+\epsilon_1+2
   \epsilon_2),
\end{split}
\end{equation}

\begin{equation}
\begin{split}
    E_2 =& -\frac{1}{2} \epsilon_1 \cos (2 \beta ) \sin (2 \theta ) (\cos (2
   \phi )-3) \\
   &+2 \epsilon_1 \sin (2 \beta ) \cos (2 \theta ) \sin
   (\phi )\\
   &+\sin (2 \theta ) (\epsilon_1+2 \epsilon_2) \cos
   ^2(\phi ),
\end{split}
\end{equation}

\begin{equation}
\begin{split}
    E_3 = 2& \cos (\phi ) [\epsilon_1 \sin (2 \beta ) \cos (\theta ) \\
    &+\sin
   (\theta ) \sin (\phi ) (\epsilon_1 \cos (2 \beta
   )-\epsilon_1-2 \epsilon_2)].
\end{split}
\end{equation}

The above expressed solution is for a droplet in stationary magnetic field pointing in the $\pvec{e}_z$ direction. 
To get the solution for a rotating field with the angular velocity $\vect{\omega}=\omega \pvec{e}_x$, we can change the reference such that the rotating magnetic field is pointing in the $\pvec{e}_z$ direction by adding a background rotating velocity field $\vect{v}_{rot}=-\omega \pvec{e}_x\times \vect{r}$ to both internal and external velocity fields.

\section{Boundary element method}
\label{sec:BEM}

\begin{figure}[h]
    \centering
    \begin{overpic}[scale=0.4]{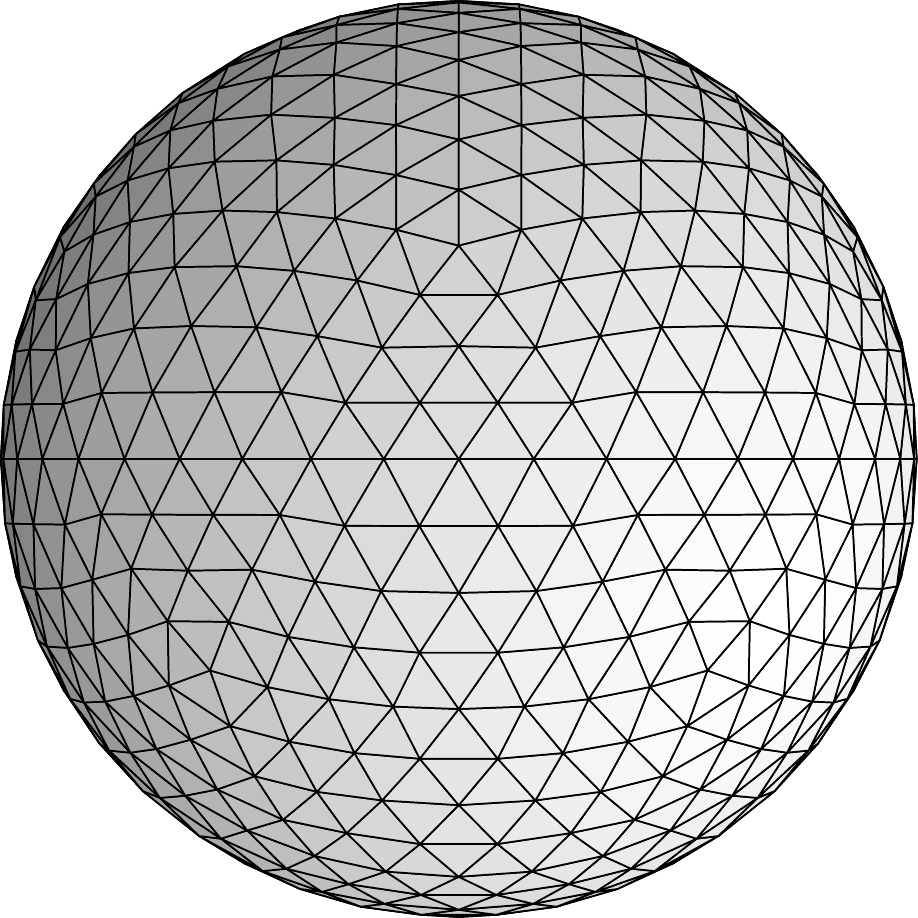}
        \put (0,100.5) {(a)}
    \end{overpic}
    \begin{overpic}[scale=0.4]{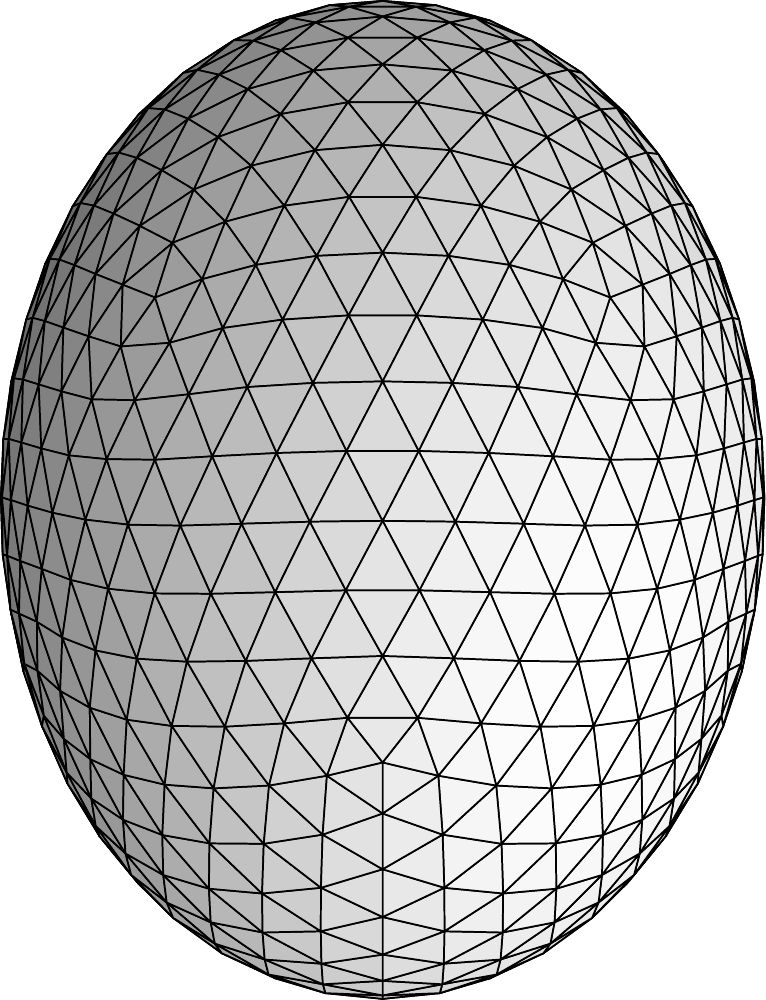}
        \put (0,92.5) {(b)}
    \end{overpic}
    \begin{overpic}[scale=0.4]{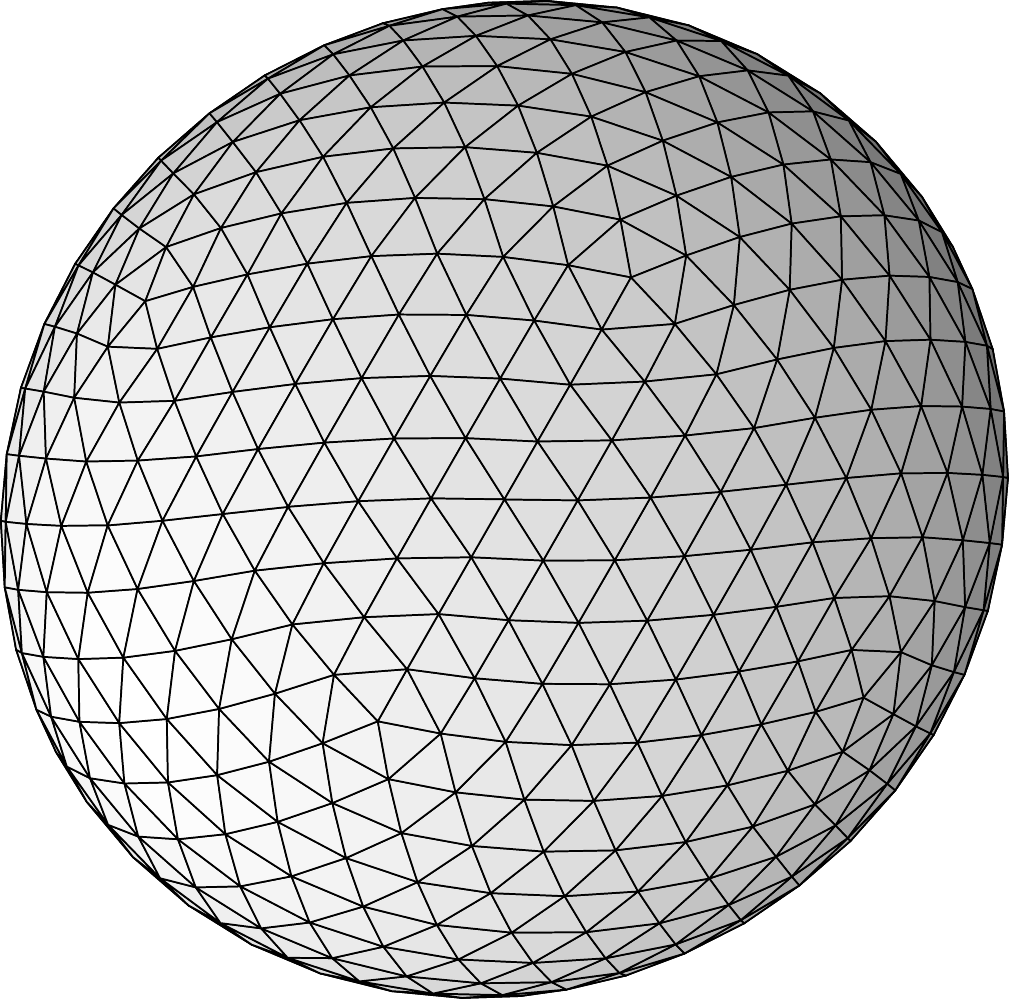}
        \put (0,87.25) {(c)}
    \end{overpic}
    \caption{The triangular mesh on the droplet's surface. (a) is the initial ($t=0$) mesh of a spherical droplet. (b) and (c) are the views of the mesh at $t=500/(R_0\ar{\eta}/\gamma)$ perpendicular and along the angular velocity of the magnetic field, respectively. The simulation parameters are $Bm=9, \mu_r=10, \lambda=100, \omega=0.05/(\gamma/(R_0\ar{\eta}))$.}
    \label{fig:mesh}
\end{figure}

The governing equations presented in section \ref{sec:problem_formulation} are written in boundary integral form. 
The Laplace equation for the magnetic potential (eq. \eqref{eq:laplacePsi}) in the integral form reads \citep{pozrikidis-practical-2002, erdmanis-magnetic-2017}
\begin{equation}
\label{eq:psi_beq}
\begin{split}
	\psi(\vect{y}) =& \frac{2\vect{y} \boldsymbol{\cdot} \vect{H}_\infty(\vect{y})}{1+\mu_r}  \\
	&+ \frac{1-\mu_r}{1+\mu_r}\frac{1}{2\pi}\int_S \psi(\vect{x}) \dd{ }{x_i}\left(\frac{1}{|\vect{X}|}\right) n_i(\vect{x})dS_x,
\end{split}
\end{equation}
where the integration is over the droplet's surface and we have introduced $\vect{X}=\vect{x}-\vect{y}$.
The Stokes equations (eq.\eqref{eq:stokes_eq}) in the integral form read \citep{pozrikidis-boundary-1992}
\begin{equation}
\label{eq:v_beq}
\begin{split}
	&v_k(\vect{y}) = \frac{2v^{\infty}_{k}(\vect{y})}{1+\lambda} \\
	& -\frac{1}{1+\lambda}\frac{\gamma}{4\pi\ar{\eta}}\int_S (k_1(\vect{x})+k_2(\vect{x}))n_i(\vect{x})G_{ik}(\vect{x},\vect{y})dS_x \\ 
	& +\frac{1}{1+\lambda}\frac{1}{4\pi\ar{\eta}}\int_S f_M(\vect{x}) n_i(\vect{x})G_{ik}(\vect{x},\vect{y})dS_x\\
	&+ \frac{1-\lambda}{1+\lambda}\frac{1}{4\pi}\int_S v_i(\vect{x})T_{ijk}(\vect{x},\vect{y})n_j(\vect{x})dS_x, 
\end{split}
\end{equation}
where $G_{ik}(\vect{x},\vect{y}) = \delta_{ij}/|\vect{X}| + X_i X_j/|\vect{X}|^3$, $T_{ijk}(\vect{x},\vect{y}) = -6 X_i X_j X_k/|\vect{X}|^5$ and $v_k^\infty(\vect{y})$ is the background flow far away from the droplet. The boundary conditions are automatically satisfied if we solve the integral equations.

For smooth droplets, all the integrands in eq. \eqref{eq:psi_beq} and eq. \eqref{eq:v_beq} scale as $1/|\vect{X}|$ as $\vec{x}\rightarrow\vec{y}$ and some steps need to be taken to facilitate their numerical evaluation. 
Details for calculating the magnetic potential (eq. \eqref{eq:psi_beq}) and from that the effective magnetic surface force can be found in the work by Erdmanis et al.\citep{erdmanis-magnetic-2017}
Some notes may be appropriate on the way we tackled the velocity integral equation (eq. \eqref{eq:v_beq}). 

To regularize the integral and avoid introducing numerical errors from calculating the curvature, the first term on the right hand side was expressed in a curvaturless form \citep{zinchenko-cusping-1999}
\begin{equation}
\begin{split}
    &\int_S(k_1(\vect{x}) + k_2(\vect{x}))n_i(\vect{x}) G_{ik}(\vect{x},\vect{y})dS_x = -\int_S \Big[\\ &
    X_i n_i(\vect{x})n_k(\vect{y})  +  X_i n_i(\vect{y})n_k(\vect{x})  +  [1 - n_i(\vect{x}) n_i(\vect{y}) ]X_k \vphantom{\frac{(num)}{(\vect{d}en)}}  \\
 &-  \frac{3 X_k  (  n_i(\vect{x}) + n_i(\vect{y}) ) X_i X_j n_j(\vect{x}) }{|\vect{X}|^2}    \Big]\frac{dS_x}{|\vect{X}|^3},
\end{split}
\end{equation}
where the expression in square brackets is proportional to $|\vect{X}|^3$ as $\vec{x}\rightarrow\vec{y}$.
The second integral can be regularized using singularity subtraction \citep{pozrikidis-boundary-1992}. We employ the identity $\int_S n_i(\vect{x}) G_{ik}(\vect x, \vect{y}) dS_x = 0$ to get
\begin{equation}
\begin{split}
    &\int_S f_M(\vect{x})  n_i(\vect{x})G_{ik}(\vect{x},\vect{y})dS_x \\
    =&\int_S [f_M(\vect{x}) - f_M(\vect{y}) ] n_i(\vect{x})G_{ik}(\vect{x},\vect{y})dS_x,
\end{split}
\end{equation}
where the integrand is now $O(1)$ as $\vec{x}\rightarrow\vec{y}$, if $f_M(\vect{x})$ is smooth.
Finally, singularity subtraction is used also on the third term \citep{pozrikidis-boundary-1992}. The identity $\int_S T_{ijk}(\vect x, \vect{y}) n_j(\vect{x}) dS_x = -4\pi\delta_{ik}$ leads to 
\begin{equation}
\begin{split}
    &\int_S v_i(\vect{x})T_{ijk}(\vect{x},\vect{y})n_j(\vect{x})dS_x\\
    =&\int_S [v_i(\vect{x}) - v_i(\vect{y})]T_{ijk}(\vect{x},\vect{y})n_j(\vect{x})dS_x - 4\pi v_k(\vect{y}),
\end{split}
\end{equation}
where the integrand is now $O(1)$ as $\vec{x}\rightarrow\vec{y}$, if $\vect v(\vect{x})$ is smooth.

We mesh the surface of the droplet with triangular elements (figure \ref{fig:mesh}) and efficiently solve the now regularized integral equations with the trapezoidal rule. 
First, we find the magnetic potential, then we compute the effective magnetic surface force $f_M$, which we use to find the velocity on the surface of the droplet. 
We move the mesh points with this velocity using the Euler method and repeat the calculation for the next time step. 
A manuscript with a full description of the numerical algorithm is in preparation.

\section*{References}
\bibliography{references}

\end{document}